# Sub-diffractional, volume-confined polaritons in a natural hyperbolic material: hexagonal boron nitride

Joshua D. Caldwell[1*], Andrey V. Kretinin[2], Yiguo Chen[3,6],
Vincenzo Giannini[3], Michael M. Fogler[4], Yan Francescato[3], Chase T. Ellis[5], Joseph G. Tischler[1],
Colin R. Woods[2], Alexander J. Giles[5], Minghui Hong[6], Kenji Watanabe[7], Takashi Taniguchi[7],
Stefan A. Maier[3] and Kostya S. Novoselov[2]

Strongly anisotropic media where the principal components of the dielectric tensor have opposite signs are called hyperbolic. Such materials exhibit unique nanophotonic properties enabled by the highly directional propagation of slow-light modes localized at deeply sub-diffractional length scales. While artificial hyperbolic metamaterials have been demonstrated, they suffer from high plasmonic losses and require complex nanofabrication, which in turn induces the size-dependent limitations on optical confinement. The low-loss, mid-infrared, *natural* hyperbolic material hexagonal boron nitride is an attractive alternative. We observe four series of multiple (up to seven) 'hyperbolic polariton' modes in boron nitride nanocones in two spectral bands. The resonant modes obey the predicted aspect ratio dependence and exhibit record-high quality factors ($Q$ up to 283) in the strong confinement regime (up to $\lambda / 86$ in the smallest structures). These observations assert hexagonal boron nitride as a promising platform for studying novel regimes of light-matter interactions and nanophotonic device engineering.

[1] U.S. Naval Research Laboratory, 4555 Overlook Ave, S.W., Washington, D.C. USA
[2] School of Physics and Astronomy, University of Manchester, Oxford Rd, Manchester, United Kingdom
[3] The Blackett Laboratory, Imperial College London, London, United Kingdom
[4] Department of Physics, University of California San Diego, 9500 Gilman Dr, La Jolla, CA USA
[5] NRC Postdoctoral Fellow (residing at NRL, Washington, D.C.)
[6] Department of Electrical and Computer Engineering, National University of Singapore, Singapore
[7] National Institute for Materials Science, 1-1 Namiki, Tsukuba, Ibaraki 305-0044, Japan
*corresponding author, E-mail: `joshua.caldwell@nrl.navy.mil`

Confining radiation to length scales much shorter than the diffraction limit at a given free-space wavelength $\lambda(\omega)$ is the primary focus of nanophotonics. There are two established approaches to this goal. One relies on materials of high positive permittivity $\text{Re}\,\varepsilon(\omega)$, wherein photons can be confined to the volume.[1-3] This allows one to achieve a high optical density of states; however, the confinement length scale cannot be smaller than $\sim \lambda/\sqrt{\varepsilon}$, which is a modest effect at visible frequencies. The other route utilizes materials where $\text{Re}\,\varepsilon(\omega)$ is negative, e.g., metals at $\omega$ below the plasma frequency.[4] In such media electromagnetic energy can be localized in the form of surface plasmons. In the absence of losses, $\text{Im}\,\varepsilon(\omega) \to 0$, the smallest wavelength of such surface excitations is limited only by the microscopic structure, e.g., the dimension $a$ of the atomic unit cell. However, in reality electronic losses are significant[5-7] making this ultimate limit difficult to attain. Hyperbolic materials[8-15] (HMs), whose permittivity tensor possesses both positive and negative principal components, have been proposed to enable novel functionalities that combine the advantages of strong confinement with full utilization of the material volume. The modes that make this possible are the high-momenta $\vec{k} = (\vec{k}_t, k_z)$ extraordinary rays – hyperbolic polaritons (HPs). The HP isofrequency surfaces are open hyperboloids, of either Type I (Fig. 1a, left) if $\text{Re}\,\varepsilon_z < 0, \text{Re}\,\varepsilon_t > 0$ or Type II (Fig. 1a, right) if $\text{Re}\,\varepsilon_z > 0, \text{Re}\,\varepsilon_t < 0$. (Here and below we assume a uniaxial anisotropy and label the tangential and axial components by $t$ and $z$, respectively.) At large $k$ the hyperboloids of both types can be approximated by cones,

$$\frac{k_z^2}{\varepsilon_t(\omega)} + \frac{k_t^2}{\varepsilon_z(\omega)} = 0, \qquad \frac{\omega}{c} \ll k_t, k_z \ll \frac{1}{a}. \qquad (1)$$

In this approximation the allowed directions for the momentum or the phase velocity of the HPs are restricted to the angle $\theta_{\text{ph}}(\omega) = \arctan\left(\frac{\sqrt{\varepsilon_z(\omega)}}{i\sqrt{\varepsilon_t(\omega)}}\right)$ with respect to the optical axis. HMs stand in a stark contrast to conventional media with positive $\text{Re}\,\varepsilon_z$ and $\text{Re}\,\varepsilon_t$ (e.g., the free space) where the magnitude rather than direction of $\vec{k}$ is restricted, resulting in a more familiar closed isofrequency surface (e.g., a sphere). The field polarization of the HP is also unusual. The electric field $\vec{E}$ is parallel to the momentum, while the magnetic $\vec{H}$ field is azimuthal. As a result, the Poynting vector $\vec{S} = \vec{E} \times \vec{H}$ is orthogonal to the phase velocity (parallel to $\vec{k}$ in Fig. 1a). Both $\vec{S}$ and the group velocity $\vec{v} = \partial \omega / \partial \vec{k}$ have a fixed angle with respect to the $z$-axis, e.g.,

$$\theta(\omega) = \frac{\pi}{2} - \arctan\left(\frac{\sqrt{\varepsilon_z(\omega)}}{i\sqrt{\varepsilon_t(\omega)}}\right) \qquad (2)$$

for the $k_z > 0$ cone. This angle, which defines the propagation direction of the HPs, will play an important role in the later discussion. The rigid directionality of HPs serves as a basis for their proposed applications in super-resolution imaging and nanolithography.[12,13]

The scale invariance of Eq. (1) implies that the HP modes can be localized inside extremely small volumes. Recent advances in nanofabrication of high-quality hexagonal boron nitride (hBN), a natural hyperbolic material, enabled the initial test of this prediction in experiments done on slabs of hBN thinned down to a thickness as small as $d_z \sim 1$ nm. These scanning near-field optical studies[16] were able to detect the HP modes with momenta $k_z \approx \frac{\pi}{d_z}$. Higher-order modes with $k_z \approx \frac{\pi}{d_z} l$, $l = 2, 3, ...$ were also predicted but not yet observed. These experiments demonstrated an ultrahigh confinement $\sim \lambda/10^4$, albeit only along the axial (*z*) direction. Experimental demonstrations of HPs strongly confined in all three directions have to this point been limited to artificial metamaterials of the Type II variety built from alternating layers of Ag and Ge.[17] However, the smallest wavelength of HPs achievable in such structures is set by the size $a$ of the repeated unit cell, below which the effective medium approximation fails.[15,18,19] This limits the use of hyperbolic metamaterials for truly nanoscale photonic applications. Indeed, the confinement reported did not exceed $\lambda/12$ in that work. Additionally, the use of metallic layers that are plagued by considerable losses[6,20] caused the resonances to be rather broad (quality factors $Q = \frac{\omega_{res}}{\Delta\omega_{res}} \sim 4$, where $\omega_{res}$ and $\Delta\omega_{res}$ are the resonant frequency and linewidth) and the number of the modes observed[17] to be limited to just the first and in some cases second orders. Alternative approaches utilizing low-loss materials, such as those incorporating polar dielectrics supporting surface phonon polaritons in place of surface plasmons can offer potential improvements and in some cases additional functionality,[21,22] such as multi-frequency hyperlensing.[22] However, in all of these previously discussed hyperbolic metamaterials, the sign of the permittivity tensor was fixed by the geometry of the hybrid structure, resulting in either Type I or Type II behavior for a given sample. This limits the design flexibility and makes it difficult to directly probe the physical differences between these two types of hyperbolic response.

In order to fully explore the fundamental phenomena and applications of HM, it is desirable to work with material systems that possess low losses, have unit cells of nanoscale dimensions, and feature both types of hyperbolic response. Here we report on experimental observations of HPs in hBN, a material satisfying all of these specifications. We detected the HP modes through optical reflection and transmission measurements performed on periodic arrays of conical hBN nanoparticles. These HP modes are manifested as resonances within two different mid-infrared spectral regions known as the lower and the upper Reststrahlen bands. We observed two distinct series of resonances per band (four in total). As many as seven modal orders in each series have been identified, consistent with the low optical losses of hBN, which are generic for polar dielectrics.[5,23,24] We have verified the fundamental hyperbolic dispersion [Eq. (1)] as the origin of the resonances by studying their dependence on polarization and the incident angle of the excitation beam, as well as on the size and shape of the nanocones, comparing these results with numerical and analytical calculations. In particular, the opposite dependence of the resonant frequency upon the modal order (ascending vs. descending in $\omega$) in the lower and upper Reststrahlen bands are the signatures of the Type I and Type II hyperbolic behavior, respectively, and so they are realized simultaneously in this material. The observed modes possess record high quality factors ($Q$ up to 283)[23] for sub-diffractional resonators and confinement ratios ($\lambda/86$ in the smallest structures). These results identify hBN as a natural HM eminently suitable for exploring new fundamental physics and represents the first foray into creating natural HM[16,25] building blocks for mid-infrared nanophotonic devices.[5]

**Results**

The mid-infrared hyperbolic response of hBN arises naturally from an extreme anisotropy of its optic phonon spectra. In general, the permittivity of a polar dielectric is negative inside its Reststrahlen band(s) and positive otherwise. The Reststrahlen bands are the spectral intervals between the longitudinal (LO) and transverse (TO) optic phonon frequencies. In hBN, a layered van der Waals crystal,[26] the *ab*- and *c*-axis-polarized phonons are widely separated in frequency so that the two Reststrahlen bands are non-overlapping. In experiment they manifest themselves as high reflectivity (low transmission) intervals in the Fourier transform infrared (FTIR) spectra. An example is shown in Fig. 1b for a 900-nm-thick hBN flake on $BaF_2$. The optical axis (our *z*-axis) of the flake is the *c*-axis of hBN (Fig. 1c, inset). The plots of the derived permittivity components Re $\varepsilon_t$ (green curve) and Re $\varepsilon_z$ (magenta curve) presented in Fig. 1c verify that Re $\varepsilon_z < 0$, Re $\varepsilon_t > 0$ in the lower band ($\lambda \approx 12.1$-$13.2\ \mu m$) while Re $\varepsilon_t < 0$ Re $\varepsilon_z > 0$ in the upper ($\lambda \approx 6.2$-$7.3\ \mu m$) band (shaded regions in Fig. 1c). This means that the lower represents a Type I and the upper band a Type II hyperbolic response.

In order to confine in three dimensions and to probe the HPs, we used electron beam lithography and reactive ion etching to fabricate conical nanoresonators from hBN flakes of thickness 80-410 nm deposited on $SiO_2$/Si, quartz, or intrinsic silicon substrates. A scanning electron microscope image of a representative array is provided as an inset to Fig. 2b. The mid-infrared reflection spectra collected from a periodic 600-nm pitch array of hBN resonators of height $d_z = $ 410 nm and diameter $d_t = 375$ nm on quartz are presented for near-normal (red curve), ~25° off-normal (green curve) and ~70° grazing incidence (blue curve) excitations (see schematic in Fig. 2a). Here and below $d_t$ stands for the diameter of a cylindrical resonator with the same height and volume as the truncated cone. In the case of the grazing incidence measurements, *p*-polarized light was used to ensure that the axial modes were efficiently excited, while the near- and off-normal excitations were performed with unpolarized light in order to couple to both axial and tangential modes. The corresponding spectra are provided in Fig. 2a and 2b, respectively. They reveal distinct series of resonances within the two Reststrahlen bands. Invariably, the stronger series of resonances within each band is excited when the external electric field $\vec{E}$ is parallel to the direction along which the permittivity is the most negative. The weaker modes are stimulated with the orthogonal polarization. For instance, within the lower band, the strongest response is observed with *p*-polarized, grazing incidence light (the blue curve in Fig. 2a), while the second set of modes is preferentially observed at near-normal excitation (the red curve with asterisks in Fig. 2a). The superposition of these resonances is observed at the off-normal excitation, resulting in Fano-like[27,28] lineshapes (the green curve in Fig. 2a). Due to the inversion of the signs of the permittivity tensors along the principal axes within the upper Reststrahlen band, the incident angle and polarization required to excite the two sets of modes are reversed (e.g. the strongest response is observed with near-normal and the weaker with p-polarized, grazing incidence excitation). Another conspicuous fact is that the two Reststrahlen bands exhibit opposite spectral trends. That is, the higher-order resonances occur at larger $\omega$ in the lower, but at smaller $\omega$ in the upper band. This behavior is common to both the strong and weak resonance series in each band.

Altogether we studied and recorded resonance frequencies for 24 different samples. Each was a periodic array of hBN nanoresonators but $d_z$ and $d_t$ were varied. Qualitatively similar results were observed from all of the arrays (see Supplementary Materials for more details); however, the positions of the resonant peaks were different for each sample. In order to test a possible scaling behavior, we have plotted these frequencies as a function of the aspect ratio

$$A = d_t/d_z. \qquad (3)$$

Remarkably, the entire data have been found to cluster along a set of well-defined smooth curves in such a plot, as shown in Fig. 3. (The labeling of the modes in Figs. 3-5 is explained in the next section.)

To investigate the physical origin of the observed resonances we carried out numerical simulations of the reflectivity spectra. The results shown in Fig. 2c and 2d theoretically verify the existence of both the strong and weak series of resonances in each Reststrahlen band, (four in total). The small discrepancies among the measured and calculated resonance frequencies may be attributed to slight irregularities in the shape of the actual nanoparticles compared to the idealized truncated cone used in the modeling. The incident polarization and angle requirements in the simulations are also consistent with those found in the experiments. Thus, the dominant resonance series within each band is excited when the external field $\vec{E}$ is parallel to the negative permittivity direction. The opposite order of the frequency shifts in the two bands (increasing in the lower Reststrahlen *vs.* decreasing in the upper Reststrahlen) is reproduced by the simulations as well.

Further insight into the nature of the modes is provided by the electric field distribution inside the particles computed for the dominant resonances in the lower and upper bands, which are presented in Fig. 4a and 4b, respectively. In an effort to uncover the impact of hyperbolicity on the field distributions, concurrent calculations were performed for fictitious isotropic materials with permittivity equal to $\varepsilon_z$ (Fig. 4c) and $\varepsilon_t$ (Fig. 4d) of hBN. The isotropic response results in surface-confined distributions typical of surface plasmons[4] and surface phonon-polaritons.[23] In contrast, in the hyperbolic case the field is volume-confined, whereby the evanescent decay of the near-fields within the volume of the hyperbolic structures is not apparent, as demonstrated in Figs. 4a and 4b. Additionally, the *E*-field distribution exhibits an intriguing 'cross-hatch' pattern, which is undoubtedly related to the directionality of the HP propagation. Indeed, the angle between the high-field crossing lines and the *z*-axis is numerically close to the angle $\theta(\omega)$ given by Eq. (2). These directional HP rays appear to be preferentially emitted by the top and bottom edges of the nanocones. The rays subsequently zigzag through the particle experiencing total internal reflections at the surface. The reason why these rays are predominantly launched at the edges is not completely clear, but we speculate it could be due to the generic tendency to find a singular field enhancement near sharp geometric features. The electric and magnetic field distributions for the principal and for several of the higher-order modes are presented in Fig. 5a and 5b. (The movies of the time-variation of these fields are provided in the Supplementary Material.) The cross-hatch pattern is evident for the stronger resonance series both in the lower (70° incidence angle, Fig. 5a) and upper (0° incidence angle, Fig. 5b) Reststrahlen bands. For these surface plots, we selected the field components along the directions where they have the highest amplitude: parallel to the negative permittivity direction for the electric field and along the azimuthal y-axis for the magnetic field. To show the magnetic field distribution due to the confined HP modes more clearly, we subtract the average $H_y$ that is largely due to the uniform illumination of the particle (the incident field plus its reflection off the underlying substrate). Higher-order modes are characterized by the progressively increasing number of the extrema (minima and maxima) of the electric (or magnetic) field.

We have also compared numerically calculated spectra of arrays with those of isolated hBN nanocones. We concluded that the inter-particle coupling in the arrays should be quite small for most of the samples studied, and so it is permissible to seek a more intuitive physical explanation of the observed spectra in terms of the mode structure of a single resonator. This is done in the next section where we address the polarization selection rules, the spectral shifts, and the origin of the data collapse as a function of the aspect ratio.

**Discussion**

Historically, exact results available for certain regular shapes, e.g., the celebrated Mie theory for the sphere, played a major role in theoretical studies of light scattering by small particles.[29] It has been known for some time that the sphere and, more generally, spheroid admits an analytical solution for the case of arbitrary uniaxial anisotropy, including the hyperbolic regime.[30,31] We found this solution to be extremely helpful for the conceptual understanding of the present case of nanocone resonators, and so we briefly discuss it below.

The eigenmodes of a spheroid with the symmetry axis parallel to the optical anisotropy axis can be labeled by three quantum numbers. The first, $m = 0, 1, ...$ is the z-axis angular momentum. The other two are the orbital index $l = m, m + 1, m + 2, ...$ and the radial index $r = 0, 1, ..., r_{max}$ (where $2r_{max} + m \leq l$) that specify the number of nodes of the scalar potential $\Phi$ (related to the electric field by $\vec{E} = -\nabla\Phi$) along the meridian and the radius. Coupling of these modes to far-field radiation is determined by the dipole selection rules. In particular, only $m = 0$ and $m = 1$ families are observable. They couple to photons polarized parallel and perpendicular to the z-axis, respectively. Actually, modes of nonzero $r$ would produce only weak resonances because of their small dipole moment. Below we neglect them and suppress the radial index assuming it is equal to zero. The observable modes of the spheroid can therefore be denoted $TM_{ml}$ where 'TM' (transverse magnetic) indicates that the magnetic field $\vec{H}$ is orthogonal to the optical axis (Figs. 1 and 2). The dipole resonances $l = 1$ are the dominant modes. They appear at frequencies satisfying the condition[29] $\varepsilon_j(\omega) = 1 - L_j^{-1}$, where $L_j(A)$ is the depolarization factor of the spheroid along the polarization direction $j = z, t$ of the incident field. Equations for higher-order modes can be found in the Supplementary Materials [see also the approximate equation (4) below]. Because of the scale invariance of the problem, the resonant frequencies depend not on the absolute dimensions of the resonator but only on its aspect ratio $A$. For example, in the $A \to 0$ limit (needle-like particle) the dipole resonance conditions $\varepsilon_z(\omega) = -\infty$ for $TM_{01}$ and $\varepsilon_t(\omega) = -1$ is for $TM_{11}$ generate the entire $TM_{0l}$ and $TM_{1l}$ series (the dominant series). In the opposite limit $A \to \infty$ (disk-like particle or a slab) the conditions to these series change to $\varepsilon_z(\omega) = 0$ and $\varepsilon_t(\omega) = -\infty$, respectively. (The permittivity of any physical system and so the solutions $\omega = \omega_{res} - i\Delta\omega_{res}$ of these equation are complex. As usual, the imaginary parts represent the broadening and the real parts are the actual resonant frequencies.)

Using the derived optical constants of hBN (Fig. 1b) we plotted the dispersion curves of the spheroid as a function of the aspect ratio in Fig. 3 (computational details are provided in the Supplementary Material), adding a left superscript (L or U) to the mode notation in order to discriminate the Reststrahlen band (lower or upper). In the lower band, the starting point $A = 0$ of the curves is at the bottom of the band, $\omega_{TO,z} = 760 \text{ cm}^{-1}$. The mode frequencies monotonically increase toward $\omega_{LO,z} = 825 \text{ cm}^{-1}$. Both these extremal points and the trend can be understood from the fact that the depolarization factor $L_z$ monotonically increases from 0 to 1 as a function of $A$. In the upper band, the curves start at $\omega = 1575 \text{ cm}^{-1}$, somewhat red-shifted from the TO phonon frequency ($\omega_{LO,t}$=1614 cm$^{-1}$), and monotonically decrease toward $\omega_{TO,t} = 1370 \text{ cm}^{-1}$. It is readily apparent that experimental data (symbols in Fig. 3) exhibit the same qualitative dependence on the aspect ratio and are even quantitatively close to the theoretical dispersion plots. The agreement is impressive given the obvious difference in shape (truncated cones vs. spheroids) and the effect of the substrate, which is disregarded in the theoretical calculations. As already mentioned, the resonance spectra depend not on the absolute dimensions of the sub-diffractional particles, but Fig. 3 also indicates that a stronger statement is valid; that the spectra are also relatively insensitive to the fine details of the particle shape as well. Taken as a guiding principle, this permits us to use the same modal notations $^R TM_{ml}$ for each of the resonances depicted in Figs. 2c, 2d, 3, and 5. This principle immediately explains the majority of our experimental observations. It elucidates the reason for the data collapse of the resonant frequencies for various sized nanostructures upon a single aspect ratio trend for a given modal order. It clarifies why the number of observed resonance series in each band is exactly two (they correspond to $m = 0$ and $m = 1$) and why the polarization selection rule for the strongest resonances exists. Indeed, such resonances correspond to the $l = 1$ dipole modes that get excited only by that electric field component $j = t, z$ along which the permittivity is negative.

A qualitative insight into the structure of weaker resonance series is provided by the semiclassical method frequently used to analyze high-order optical modes $l, m \gg 1$ (e.g., whispering-gallery modes) of dielectric resonators. In the semiclassical picture the HPs propagate inside the nanoresonator as directional rays that experience total internal reflections at the boundaries. Depending on the shape of the particle, the mechanics of such rays can be either chaotic or regular.[32] In the spheroidal geometry we have the latter, in which case we have a simple quantization rule. It is basically the standing-wave conditions $k_z d_z \sim l$ and $k_t d_t \sim m$ imposed on the characteristic momenta $k_z$ and $k_t$ of the HP. We can combine these conditions with Eqs. (1) and (2) to find the equation for the eigenfrequencies $\omega$:

$$A \sim \frac{m}{l} \tan \theta(\omega) = \frac{m}{l} \frac{i\sqrt{\varepsilon_t(\omega)}}{\sqrt{\varepsilon_z(\omega)}}. \tag{4}$$

The qualitative trends of the spectra can be understood without expressly solving Eq. (4). In the lower Reststrahlen band the right-hand side of Eq. (4) is the increasing function of $\omega$. Hence, if $m$ and the aspect ratio are fixed, the resonance frequency should be higher for higher index $l$. In the upper Reststrahlen band, the right-hand side of Eq. (4) is a decreasing function, and so the modal order is opposite. This predicted inversion of these trends is in agreement with our observations (Fig. 2). By a similar argument, the frequencies of different modes $^R\text{TM}_{ml}$ should increase (decrease) as a function of the aspect ratio for R = L (U). The anomalous behavior of the upper Reststrahlen band was also reported within Ag/Ge Type II hyperbolic metamaterial resonators.[17] Note that the spectral shifts have the same sign as the $z$-component of the group velocity $v_z = \partial \omega / \partial k_z$ of the HP. As shown by the green arrows in Fig. 1a left (right), $v_z$ is positive (negative) in Type I (II) case.

Extrapolated to the case $m = 1$, Eq. (4) implies a simple rule regarding the discussed cross-hatch pattern of the electric field distribution. From elementary geometrical considerations, the total number of the ray intersections along the nanocone axis should be equal to $N = A^{-1} \tan \theta(\omega)$. In view of Eq. (4), this yields $N \sim l/m = l$, i.e., the number of intersections is equal to the modal index. This approximate rule is in fact closely obeyed by all the $m = 1$ modes shown in Fig. 5 (right). Although no such simple rule can be given for $m = 0$, we may expect $N > l$ in this case, which is also borne out by the simulations (Fig. 5, left).

Concluding this discussion, it is instructive to compare the figures-of-merit of our nanoresonators with those in literature. The modes observed here exhibit narrow linewidths and high quality factors: $Q_{\text{tot}}$ =156-264 and 66-283, respectively, in the Type I lower and Type II upper Reststrahlen bands. These $Q$-factors are almost two orders of magnitude higher than $Q_{\text{tot}} \sim$ 4 reported for Type II hyperbolic metamaterial nanostructures,[17] and corresponds to record values for sub-diffractional modes.[23,33,34] As the $Q$-factor defines the energy stored to energy loss rate, this can be inferred as a metric defining the resonator efficiency. These high quality factors are also coupled with exceptional modal confinements as the size of our nanoresonators are 17-86 times smaller than the free-space wavelength in the $z$-direction and 6-61 times smaller in the transverse direction. Such mode confinement factors are unprecedented in metal-based surface-plasmon nanostructures. For comparison, the highest confinement achieved for hyperbolic metamaterial structures was $\lambda/12$.[17] While similar confinements have been reported for plasmons in patterned graphene,[35,36] strong electronic losses limited the $Q_{tot}$ to values between 2.5 and 15. Such losses are fundamentally absent in hBN due to its polar dielectric nature.[5,23,24]

In summary, this work constitutes the first experimental observation of sub-diffractional guided waves confined in all three dimensions inside a natural hyperbolic material. Compared to similar efforts in metamaterials,[17] the results presented here demonstrated not only Type II but also Type I modes, which enabled the first exploration such confinement in a Type I hyperbolic medium and provided the experimental investigation of their reciprocal spectral behaviors. Weaker resonances related to theoretically predicted[16] higher-order HP modes have been identified experimentally in both Reststrahlen bands, and a more complete mode classification has been proposed. Both Reststrahlen bands supported two distinct series of HPs, which exhibited orthogonal requirements on the incident angle and polarization. This is the first description and experimental observation of these various series of HP modes. Future work may pursue the realization of hBN resonators of atomic dimensions[19,37,38] in order to achieve an ultimate confinement of these low-loss electromagnetic modes. Investigations into other polar van der Waal's materials[26] such as $MoS_2$, may further extend the spectral range where natural HMs are found.[25,39] These efforts may lead to the development of disruptive technologies such as the nanoscale equivalent of an optical fiber, due to the volume-bound confinement of sub-diffraction modes within hBN. Impact in areas such as enhanced molecular spectroscopy,[40] nanophotonic circuits,[37] tailored thermal emission sources,[41,42] flat optics,[43] and super-resolution imaging[12,44] is also anticipated.

**Acknowledgements**

Funding for NRL authors was provided by the NRL Nanoscience Institute. J.D.C. carried out this work through the NRL Long-Term Training (Sabbatical) Program at the University of Manchester. C.T.E. and A.J.G. acknowledges support from the NRC NRL Postdoctoral Fellowship Program. Y.F., S.M. and V.G. acknowledge support from EPSRC and Leverhulme Trust. C.R.W., A.K. and K.S.N. acknowledge support from the Engineering and Physical Sciences Research Council (UK), The Royal Society (UK), European Research Council and EC-FET European Graphene Flagship. M.M.F. acknowledges support from the Office of Naval Research and the University of California Office of the President. J.D.C. would like to thank Dr. Jeremy Robinson and Dr. Igor Vurgaftman of NRL for helpful discussions and comments.

**Methods**

Flakes of hBN were exfoliated from crystals grown via the high pressure/high temperature method[45,46] and deposited on either 90-nm thick $SiO_2$ on doped *n*-type Silicon, quartz or instrinsic silicon substrates using standard dicing tape. Using atomic force microscopy, the flakes of thicknesses 60-410 nm and spatial sizes sufficient for nanostructure array patterning (>50 $\mu$m on a side), were identified. The array patterns were defined using electron beam lithography and were patterned into a bilayer PMMA resist, with nanostructures ranging from 100-1000 nm in diameter being fabricated, with the periodicity (center-to-center) being maintained at a 3:1 ratio with the designed diameter. A 30-nm thick aluminum film was deposited using thermal evaporation for the standard liftoff procedure and the hBN nanostructures were created by reactive ion etching in an inductively coupled $CHF_3/O_2$ plasma (Oxford Instruments PlasmaLab) at a pressure of 10 mTorr, resulting in an etch rate of ~1$\mu$m/min. The RF power applied to the plasma reactor and the substrate was 200 W and 20 W, respectively. The $CHF_3/O_2$ flow ratio (17.5 / 10 sccm) was optimized to increase the BN-to-$SiO_2$ selectivity. The etch rates of the aluminum hard mask and $SiO_2$ substrates were determined to be insufficient for any significant etching to occur within this time. The aluminum hard mask was then removed via a standard TMAH (MICROPOSIT® MF-319) wet chemical etchant.

Mid-infrared reflectance measurements were undertaken using a Bruker Hyperion microscope coupled to a Bruker Vertex80v FTIR spectrometer. Three different objectives were used to obtain normal (ZnSe), off-normal (36x Cassegrain) and grazing (15x grazing incidence objective, GAO) incidence spectra from the nanostructures. The spectra were collected with a 2 cm$^{-1}$ spectral resolution and spatial resolution defined by the internal adjustable aperture of the microscope that was set to the size of the specific array of interest. All measurements were performed in reference to a gold film.

To better understand the deeply sub-diffractional HP modes involved in these nanostructures, full-wave electrodynamic calculations were performed using the RF module of the finite-element package COMSOL. The anisotropic optical constants were extracted from the above mentioned method and the geometry of the experimental nanostructures was determined from atomic force microscopy and scanning electron microscopy. The simulations were undertaken for the light incident at 0°, 25° and 70° angles to match the illumination conditions of the ZnSe (0.08 NA), 36x Cassegrain (0.52 NA) and GAO objectives of the Bruker FTIR microscopes described above.

**Figure Captions:**

**TOC Fig:** We observe deeply sub-diffractional low-loss hyperbolic polariton resonances which permit nanoscale photonic confinement and highly directional propagation through the volume of the natural hyperbolic material hexagonal boron nitride. In contrast to hyperbolic metamaterials, both Type I and II behavior is realized in the same sample.

**Fig. 1: Behavior of the Natural Hyperbolic Material hexagonal Boron Nitride.** **a** Schematic isofrequency surfaces for a Type I (left) and II (right) hyperbolic media. The arrows indicate the directions of the electric and magnetic fields, the Poynting vector, the momentum, and the group velocity of the HP modes. The isofrequency contours each correspond to different spectral regions of the hBN **b** FTIR reflection (red curve) and transmission (blue curve) spectra of a 900-nm thick, hBN flake exfoliated onto a 500-µm thick $BaF_2$ substrate. The frequencies of the transverse (TO) and longitudinal (LO) optical phonons of hBN in the axial and tangential directions are shown by the vertical dashed lines. A schematic of the hBN crystal structure is presented in the inset. **c** Derived real parts of the permittivity tensor components of hBN. The Type I lower and II upper Reststrahlen bands are shaded.

**Fig. 2: Hyperbolic Polaritons in hBN Nanocones.** FTIR reflection spectra of the **a** lower and **b** upper Reststrahlen bands collected from a periodic array (600 nm center-to-center pitch) of 375-nm diameter, 410-nm tall hBN conical nanoresonators. The spectra were collected using three different IR objectives giving rise to incident angles near-normal (<2°, red curve), off-normal (~25°, green curve) and at near-grazing incidence (~70°, blue curve). For clarity, the near-grazing incidence spectra is offset by -4% in **b**. Inset: a SEM image of a representative hBN nanoresonators array. **c** and **d** The spectra calculated for both Reststrahlen bands using the finite element method. The mode assignments are explained in the text.

**Fig. 3: Aspect Ratio Dependence of HP Resonances.** Analytical calculation of the aspect ratio dependence of the resonant frequencies of spheroidal nanoresonators in the **a,** upper and **b,** lower Reststrahlen bands. For simplicity, only $^{L}TM_{0l}$ and $^{U}TM_{1l}$ modes are shown. The measured resonance frequencies for the hBN conical nanoresonators are shown by symbols. Different symbols refer to data obtained for arrays with different nanoresonator heights. In all, the data was collected from 24 different arrays of varying heights and diameters.

**Fig 4: Comparison of Hyperbolic and Isotropic Behavior for hBN.** Simulated reflection spectra for the lower (left column) and upper (right column) Reststrahlen bands at 70° and 0° incident angles **a,b** Case of a hyperbolic material with anisotropic optical constants (top) **c,** Same for a fictitious isotropic material with permittivity $\varepsilon_t$ **d,** Same for a material with permittivity $\varepsilon_z$. The spatial distributions of $|E|$ for the highest amplitude peak in each graph are provided as insets.

**Fig. 5: Electromagnetic Field Simulations of Type I and II HPs in hBN nanocones.** Simulated surface plots of the components of the electric (*E*, left column) and magnetic (*H*, right column) fields for each of the resonant modes observed within the **a,** lower and **b,** upper Reststrahlen bands at 70° and 0° incident angles, respectively. The assignment of the modes is shown in Fig. 2c and 2d.

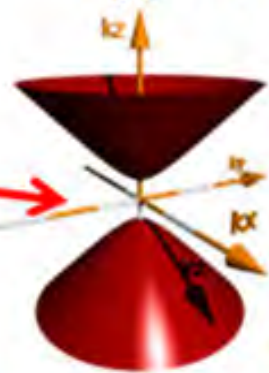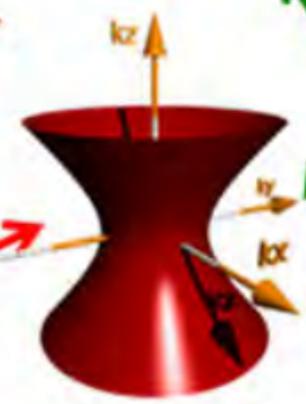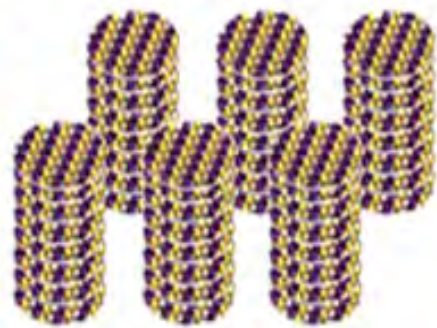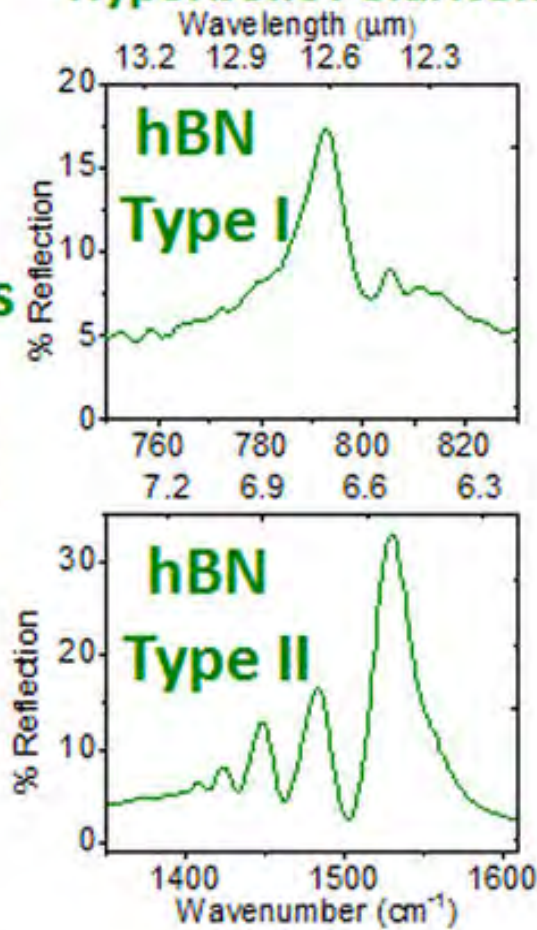

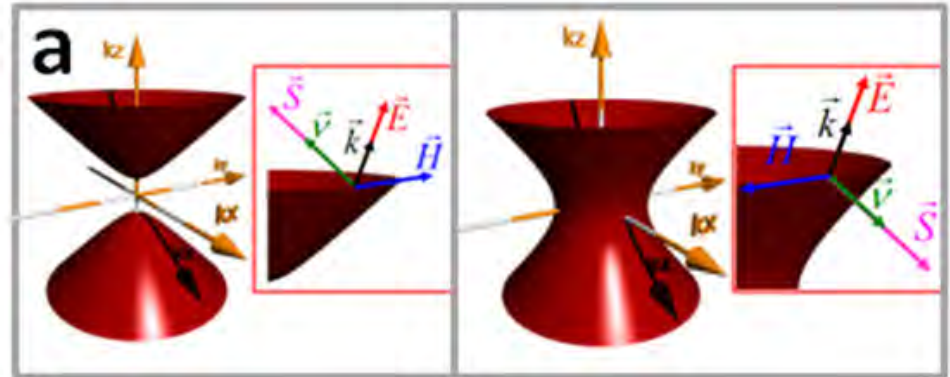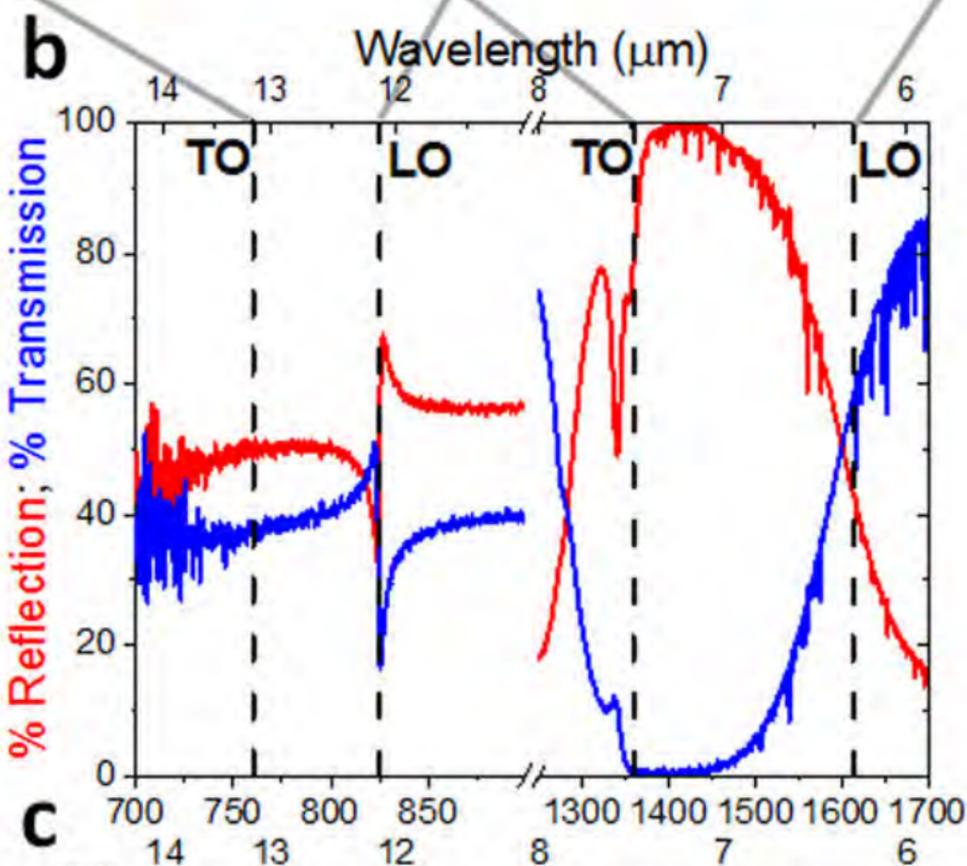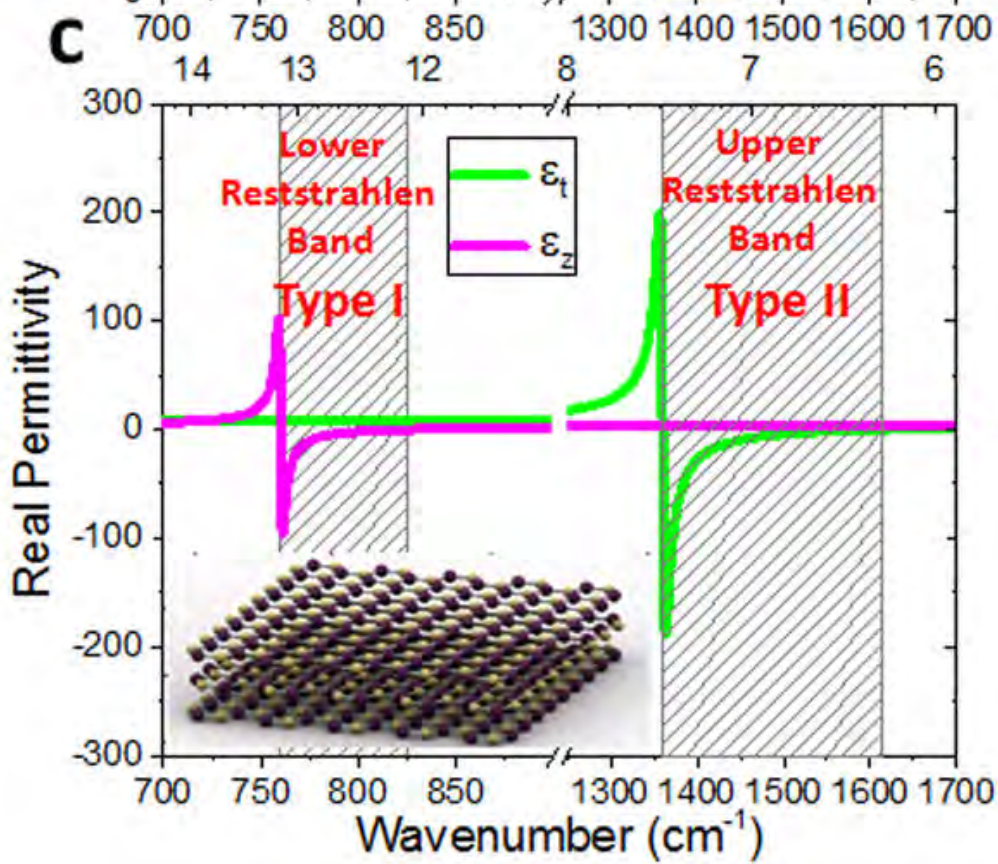

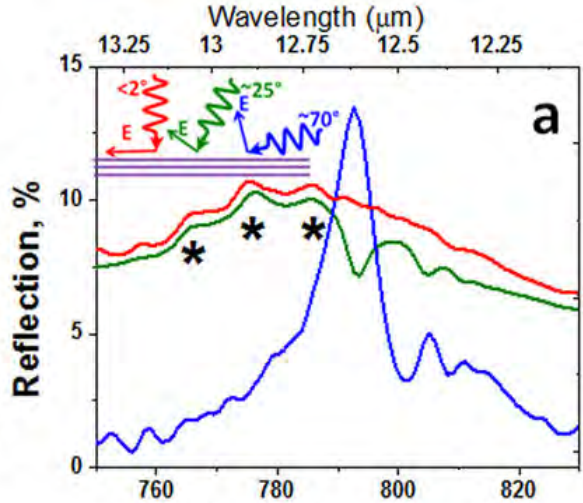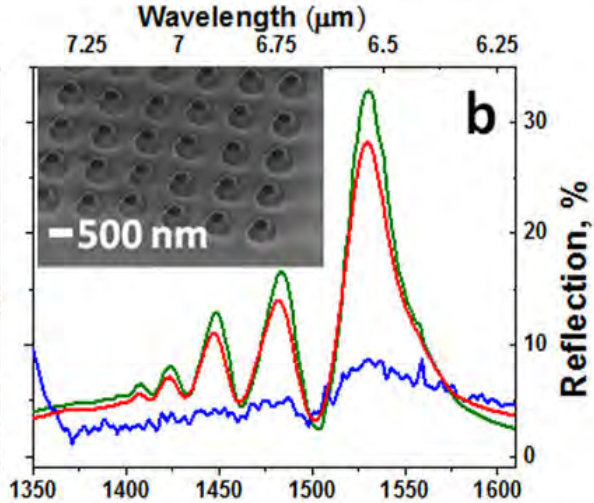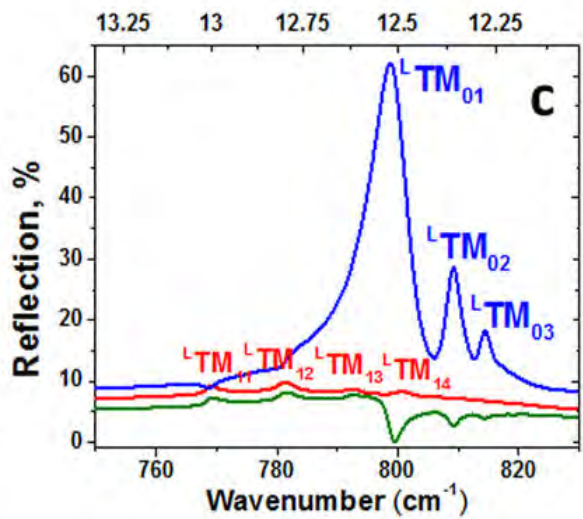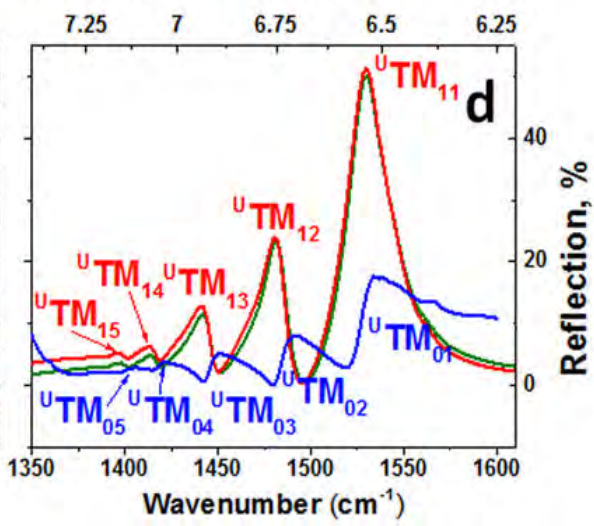

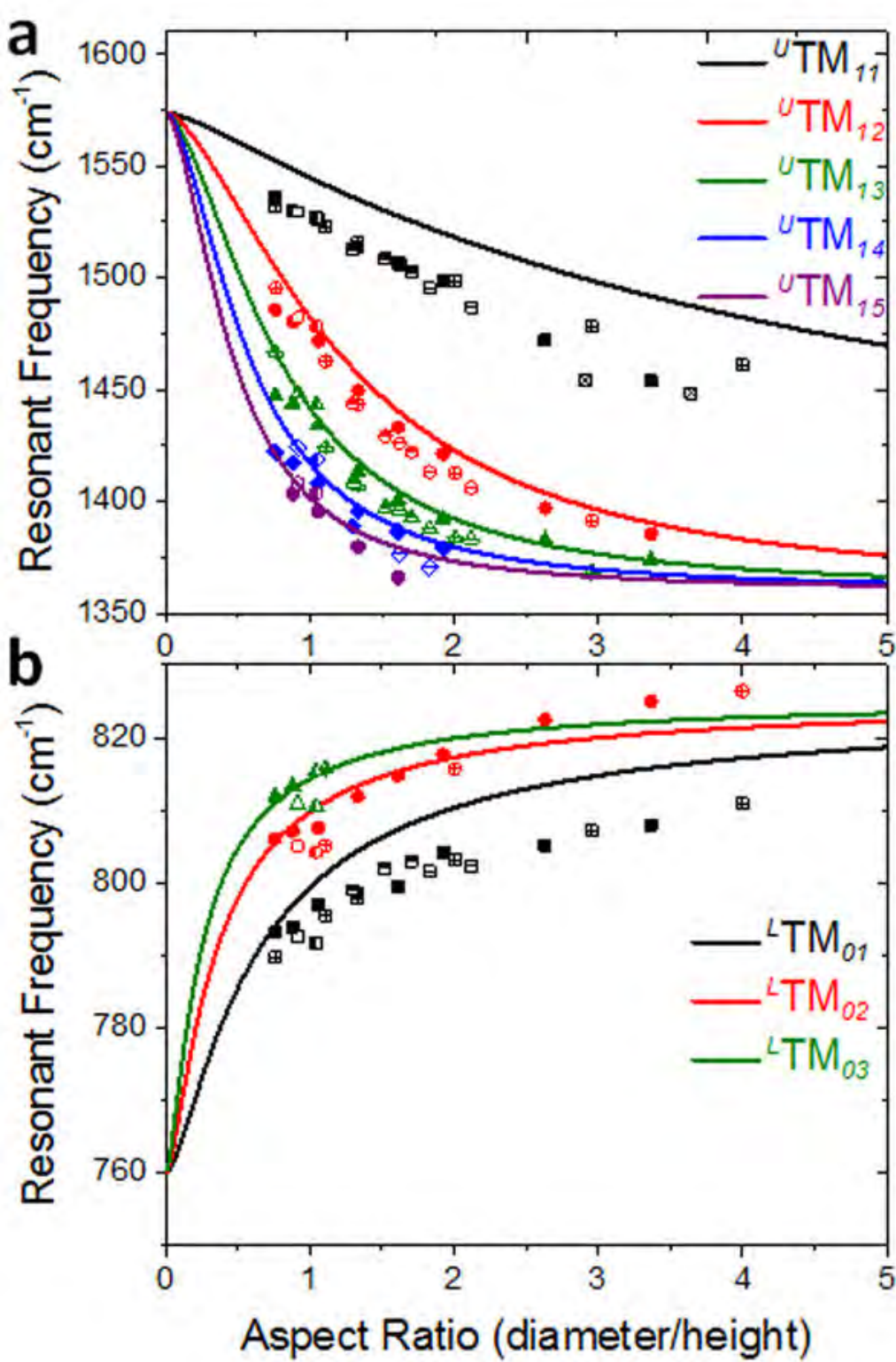

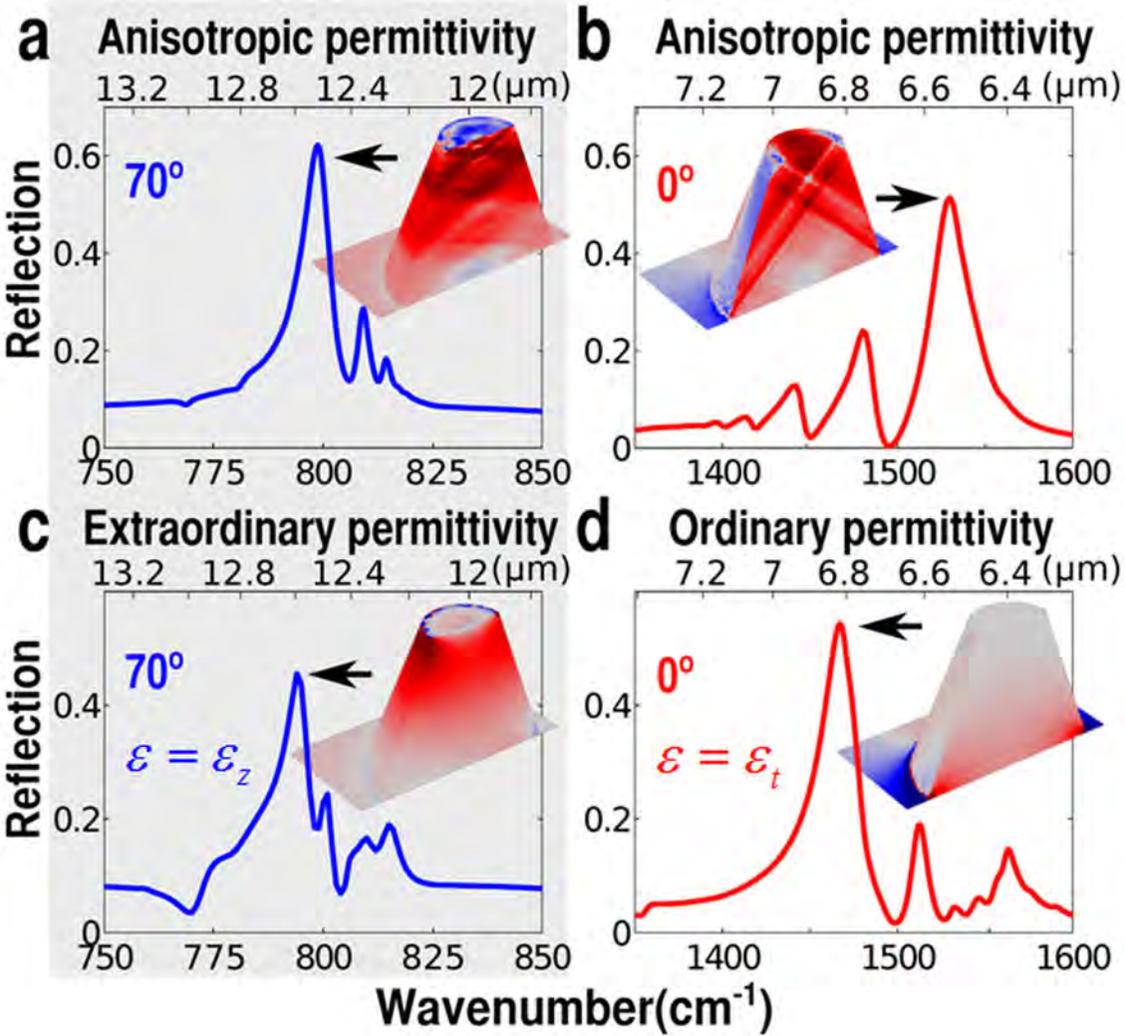

## Lower band

**a** Color: $E_z$   Color: $H_y$

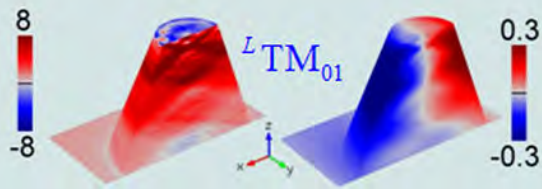

$^L TM_{01}$

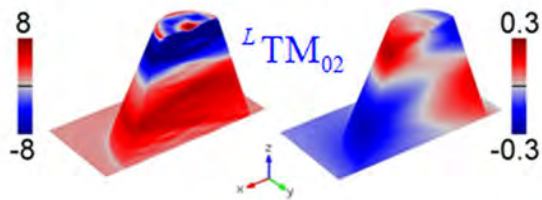

$^L TM_{02}$

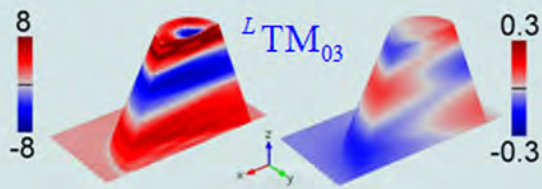

$^L TM_{03}$

## Upper band

**b** Color: $E_x$   Color: $H_y$

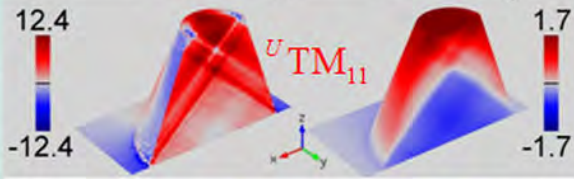

$^U TM_{11}$

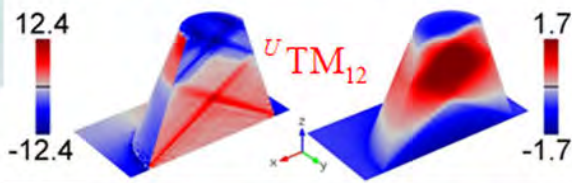

$^U TM_{12}$

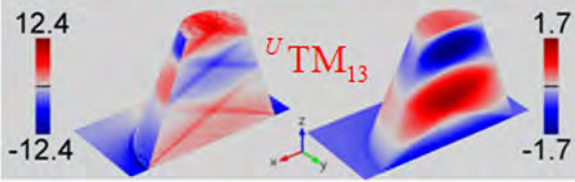

$^U TM_{13}$

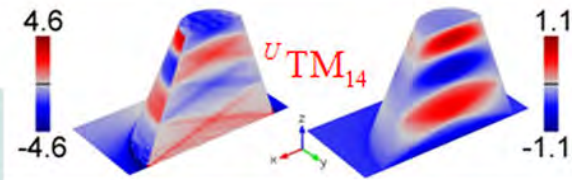

$^U TM_{14}$

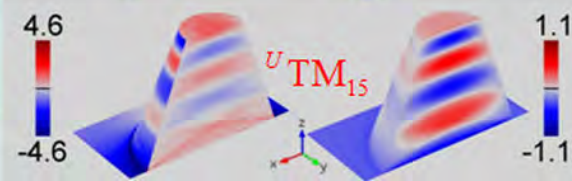

$^U TM_{15}$

*Sub-diffraction, Volume-confined Polaritons in the Natural Hyperbolic Material, Hexagonal Boron Nitride*

**Joshua D. Caldwell[1], Andrey Kretinin[2], Yiguo Chen[3,6], Vincenzo Giannini[3], Michael M. Fogler[4], Yan Francescato[3], Chase T. Ellis[5], Joseph G. Tischler[1], Colin R. Woods[2], Alexander J. Giles,[5] Minghui Hong,[6] Kenji Watanabe,[7] Takashi Taniguchi,[7] Stefan A. Maier[3] and Kostya S. Novoselov[2]**

[1] U.S. Naval Research Laboratory, 4555 Overlook Ave, S.W., Washington, D.C. USA
[2] School of Physics and Astronomy, University of Manchester, Oxford Rd, Manchester, United Kingdom
[3] The Blackett Laboratory, Imperial College London, London, United Kingdom
[4] Department of Physics, University of California San Diego, 9500 Gilman Dr, La Jolla, CA USA
[5] NRC Postdoctoral Fellow (residing at NRL, Washington, D.C.)
[6] Department of Electrical and Computer Engineering, National University of Singapore, Singapore
[7] National Institute for Materials Science, 1-1 Namiki, Tsukuba, Ibaraki 305-0044, Japan
*corresponding author, E-mail: joshua.caldwell@nrl.navy.mil

**Hexagonal Boron Nitride Optical Constants.** To derive the optical constants used in the electromagnetic modeling presented in Fig. 2, 4 and 5 and the analytical calculations in Fig. 3 of the main text, fitting of the reflection and transmission spectra from several hexagonal boron nitride (hBN) flakes that were exfoliated onto BaF$_2$ substrates were undertaken. The spectra were fitted to calculations based on the standard transfer matrix approach, assuming the single-Lorentzian form of the principal dielectric tensor components[1-3]

$$\varepsilon^a(\omega) = \varepsilon_\infty^a \left( 1 + \frac{\left(\omega_{LO}^a\right)^2 - \left(\omega_{TO}^a\right)^2}{\left(\omega_{TO}^a\right)^2 - \omega^2 - i\omega\gamma^a} \right) \quad (1)$$

where $a$ refers to either the transverse (a,b crystal plane) or z- (c crystal axis) axes, parameters $\omega_{LO}$, $\omega_{TO}$, $\varepsilon_\infty$ and $\gamma$ correspond to the LO and TO phonon frequencies, the high-frequency dielectric permittivity and the damping constant, respectively. The initial guess for these parameters was taken from Cai et al.,[4] while the BaF$_2$ optical response was taken from the optical constants provided in Palik's compendium.[5] Subsequent comparisons between the simulated and experimental (Fig. 1a) reflection spectra for a 900 nm thick hBN flake on a 500 μm thick BaF$_2$ substrate were undertaken. The parameters derived from the best-fit are provided in Table S1. They are consistent with the literature.[4,6,7] The calculated and the measured reflection spectra are shown in Fig. S1 and they demonstrate the quantitative agreement.

The derived real part of the permittivity was provided in Fig. 1c of the main text and is provided again, along with the imaginary component in Fig. S2. As stressed in the main text, hBN behaves as a natural hyperbolic material in the infrared. It exhibits negative values of the dielectric function along the transverse (ordinary) and z- (extraordinary) axis in the upper and lower Reststrahlen bands, respectively, while the orthogonal principal axis has a dielectric function with a positive real part. This anomalous optical behavior is accompanied by a relatively small imaginary part of the dielectric tensor (Fig. S2b) compared to plasmonic materials, thus implying lower dissipative losses. For instance, at the resonance condition for a sphere Re $\varepsilon(\omega) = -2$, Im $\varepsilon(\omega) \cong 0.1$ for both Reststrahlen bands of hBN, which is a factor of 3-6 times smaller than the values reported for silver,[5,8] the best plasmonic material, at the same resonant condition. Further comparisons between hBN and other polar dielectrics with plasmonic media can be found in a recent review.[9]

**Table S1:** Parameters derived from fitting the reflection spectra in Fig. 2a to transfer matrix calculations. See also Fig. 2b and c of the main text.

| Optical axis | Spatial axis | $\omega_{TO}$ (cm$^{-1}$) | $\omega_{LO}$ (cm$^{-1}$) | $\varepsilon_0$ | $\varepsilon_\infty$ | $\gamma$ (cm$^{-1}$) |
|---|---|---|---|---|---|---|
| c | z | 1360 | 1614 | 6.90 | 4.90 | 7 |
| a and b | x and y | 760 | 825 | 3.48 | 2.95 | 2 |

As discussed in the main text, the lower (~12.1-13.2 µm) and upper (~6.2-7.3 µm) Reststrahlen bands of hBN correspond to the Type I and II hyperbolic response, respectively. Hyperbolic media are characterized by the isofrequency surfaces in the wavevector (*k*-vector) space that are modified from conventional (closed) spherical or ellipsoidal shapes to hyperboloids,[10-13] see Fig. S3a-c. The Type I and II designation refers to the number of principal axes supporting negative permittivity. In an isotropic media (Fig. S3a), the closed-shell, spherical isofrequency curve enables resonant excitations with a *k*-vector similar to that for free-space $(k = \omega/c)$, regardless of incidence angle. Modes with *k* exceeding this upper limit are evanescent. In the case of hyperbolic materials, the length of the *k*-vector can be arbitrarily large and therefore the modal volume (photonic confinement) arbitrarily small, while the mode can still propagate through the volume of the material, rather than exhibit an evanescent decay. This has many implications, e.g., hyper- and superlensing.[14,15]

Further reflection spectra were collected for hBN flakes of varying thicknesses ranging from 60-6400 nm (Fig. S4a) and the results agreed well with calculated spectra and the corresponding trends with increasing thickness (Fig. S4b). The positions of the longitudinal (LO) and transverse (TO) optic phonons, which define the upper and lower frequency limits of the Reststrahlen band are labeled for the upper Reststrahlen region. As the thickness of the hBN flakes is reduced, the high reflectivity upper Reststrahlen band narrows significantly and red-shifts, resulting in a band centered at the position of the TO phonon (~1360 cm$^{-1}$) for flake thicknesses <100 nm. This is consistent with an increased damping of the LO phonon with reduced hBN thickness. As the LO phonon is an out-of-plane vibration, high fidelity of this collective mode should be required to induce the high reflectivity, high frequency edge of the upper Reststrahlen band. Thus, the increased LO phonon damping with reduced thickness presumably causes this shrinking Reststrahlen band. This effect is clearly demonstrated in the calculated hBN reflection spectra in Fig. S4b and therefore demonstrates that the overall physical phenomena governing this effect is encompassed within the basic physical understanding of the system.

**Characterization of hBN Conical Nanoresonators.** The geometry of the conical nanostructures was extracted from both AFM measurements of the top surface of the hBN nanostructures and through the use of SEM imaging. As demonstrated in the SEM provided as Fig. S5, a small number of the nanostructures detached from the substrate surface during the fabrication process, thereby providing an opportunity to directly measure the top $(d_{top})$ and bottom $(d_{bottom})$ diameters and side-wall slope angles for a variety of the nanoresonators. To characterize the transverse dimension of these conical structures by a single number, we use the diameter as quantified by the volume-preserving method:

$$d_{avg} = \sqrt{\frac{1}{3}\left(d_{bottom}^2 + d_{top}^2 + d_{bottom}*d_{top}\right)^{1/2}} \tag{2}$$

This method provides the corresponding diameter for a cylinder of the same height and volume. For example, the nanostructure presented in Fig. S5 corresponds to a $d_{avg}$ = 360 nm. Linear confinements were determined by taking the ratio of the free-space wavelength of the mode at the frequency of the primary $^U\text{TM}_{11}$ and $^L\text{TM}_{01}$ resonances with the nanoresonator height $h$ and the average diameter $d_{avg}$.

As discussed in the main text, 24 separate arrays of hBN nanoresonators were fabricated and FTIR reflection spectra were collected from each. In the case of the grazing incidence objective, the reported reflection spectra correspond to the square root of the measured value. This is due to the orientation of the objective, which results in two reflections on the sample surface prior to collection of the reflected signal. This multiple path process enables the maintenance of the incident polarization, however, accomplishes this at the expense of a reduction in the collected intensity on the order of the square of the reflection. All grazing incidence spectra provided in the main text and in the Supplemental Materials have therefore been corrected accordingly.

The periodic arrays of the conical hBN nanoresonators each supported four series of resonant HP modes, two per Reststrahlen band. Within a given band, the two series exhibited orthogonal polarization and incident angle selection rules and each series of modes had an intense, 1st order resonance and several higher order resonances found at lower (higher) frequencies for the upper (lower) Reststrahlen bands. As noted in the text, this anomalous, reciprocal behavior of the spectral shift for higher order modes can be accounted for by the inversion of the sign of the *z*-component of the group velocity of the two types of hyperbolic response. In the case of hyperbolic media, for a given wavevector, the corresponding group velocity $v = (v_x, v_y, v_z)$ vector is normal to the isofrequency surface (see green arrows in Fig. S3b and c). For deeply sub-diffractional waves relevant here, $v$ and $k$ vectors are nearly orthogonal to each other, unlike the case in isotropic media where they are parallel. Based on the optical constants for hBN presented in Fig. S2a and b, the upper Type II (lower Type I) Reststrahlen band, $v_z$ is negative (positive) due to the corresponding concave (convex) shape of the isofrequency surfaces. In a bulk material, the sign of $v_z$ determines how the modal frequency changes as a function of $k_z$ wavenumber while the in-plane wave numbers $k_x$, $k_y$ are held constant. Furthermore, for sub-diffractional waves $v_z$ become a function of the dimensionless numbers $k_z / k_x$ and $k_z / k_y$. As mentioned in the main text, in a finite-size nanoresonator, each mode can also be assigned $k_z / k_x = k_z / k_y$ proportional to the aspect ratio. Therefore, the increase or decrease of the resonance frequency with the aspect ratio is also determined by the sign of $v_z$. Indeed, as demonstrated in Fig. S6 a for the $^U\text{TM}_{1l}$ resonances, for a fixed height nanostructure, increasing the diameter induced a red-shift in the resonant position. A red shift is also observed with decreasing height (thickness) (Fig. S6 b). The overall trends for the 1st order modes are also replicated for each of the higher order resonances and the relevant trends for the lowest order resonances are provided in Fig. S7a and b as a function of diameter and height, respectively, for each of the nanoresonator arrays studied. Note that for hBN nanoresonator arrays fabricated on SiO$_2$/Si substrates that the detection of the higher order Type I resonances within the lower Reststrahlen band was somewhat restricted due to the high absorption of the underlying, highly doped silicon.

In an effort to differentiate absorption from scattering effects, a large area array of 375 nm diameter, 360 nm tall nanoresonators on a 600 nm pitch were fabricated on an intrinsic silicon substrate. Due to the lack of free carriers within the intrinsic silicon, this substrate provides a transparent material over the entire spectral range extending from the lower to the upper Reststrahlen bands of hBN, and thus enabled the collection of both transmission and reflection spectra from the nanoresonator arrays, which are provided in Fig. S8. Both spectra were referenced to the background silicon substrate. Due to the ~25° incident angle associated with the 36X objective, a superposition of the $^{L}TM_{0l}$ and $^{L}TM_{1l}$ modes are observed in both spectra, with the latter modes designated with an '*'. This is consistent with the observations reported in the main text in Fig. 2a. However, the $^{L}TM_{0l}$ modes are observed as 'dips' in both reflection and transmission, indicating that these are predominantly forward scattered into the substrate or absorbed. In contrast, the $^{L}TM_{1l}$ as well as the $^{U}TM_{1l}$ modes are both observed as 'dips' in transmission, but as 'peaks' in reflection, indicating that these spectral features are predominantly dependent upon back-scattering. It is interesting to note that both $^{L}TM_{1l}$ and $^{U}TM_{1l}$ resonances are stimulated with in-plane polarization and both exhibit similar scattering phenomena, despite being observed in different Reststrahlen bands with reciprocal hyperbolic behavior.

From the least squares fitting of the various resonances, the Lorentzian linewidths were also extracted and the corresponding quality factors were derived using the formula:

$$Q = \frac{\omega_{res}}{\Delta\omega_{res}} \qquad (3)$$

with $\omega_{res}$ and $\Delta\omega_{res}$ corresponding to the resonant frequency and linewidth, respectively. As the *Q*-factor provides the ratio of the energy stored to the energy loss-rate, this provides a figure of merit for comparing resonators of various sizes, materials and resonant frequencies. The *Q*-factors extracted from two samples, a 410 nm tall on quartz (blue squares) and a 360 nm tall on intrinsic Si (red circles) are provided in Fig. S9 a for the first six order resonances in the upper ($^{U}TM_{1l}$; closed symbols) and the first 4 in the lower ($^{L}TM_{0l}$; open symbols) Reststrahlen bands. The values for these two bands ranged from 66-283 and 156-264, respectively, with the values increasing with increasing modal order. As noted in the text, these values constitute record high values for sub-diffraction optical modes, well exceeding the theoretical maximum for plasmonics (Ag spheres)[16] and the highest values for SPhP resonators that were observed in SiC nanopillars reported by several of the authors of this work.[17] Quality factors extracted from the other hBN nanostructures explored in this work were of similar magnitudes and demonstrated similar trends. The corresponding resonant frequencies are provided in Fig. S9 b and demonstrate that the resonant frequencies are relatively insensitive to the substrate index of refraction, with similar trends in the resonant response observed from nanoresonator arrays on SiO$_2$ $(n_{sub} \sim 2.5)$ and Si $(n_{sub} \sim 3.4)$. Note that the reported indices of refraction for the two materials are average values within the two Restsrahlen bands of hBN and thus do not reflect the corresponding Reststrahlen band of $SiO_2$ found between the two bands of hBN.

**Finite Element Calculations.** To better understand the deeply sub-diffraction, hyperbolic optical modes involved in these highly dispersive nanostructures, full-wave, 3D electrodynamic calculations using the RF module of the finite-element package COMSOL were performed. In an effort to closely match the experimental conditions, the simulations were undertaken using incident optical fields at 0°, 25° and 70° to match that of the ZnSe (0.08 NA), 36x Cassegrain (0.52 NA) and grazing incidence (GAO) objectives of the Bruker FTIR microscopes used.

Simulated reflectance spectra for both 0° and 70° incident angle are provided in Fig. S10 a and b for the lower and upper Reststrahlen bands, respectively. As described in the main text, two sets of resonant modes are predicted within each Reststrahlen band, which were observed experimentally. Within a given Reststrahlen band, the two sets of modes exhibit orthogonal polarization dependences, with the highest amplitude response observed when the incident electric field is aligned along the principal axis with negative real permittivity.

The overall assignment of the modes can be derived from the $|H|$ and $|E|$ near-field distributions for each resonant mode. These profiles are provided in Figs. S11 and S12 for each of the observed $^L\text{TM}_{0l}$ and $^U\text{TM}_{1l}$ resonances labeled in Fig. S10 a and b. The modal order is derived by the number of oscillations in the phase of the $H$ field, while the number of crossing $E$ field lines also increases with the modal order. Videos depicting the time-variant $H$ and $E$ near fields for the 3$^{rd}$ order $^L\text{TM}_{0l}$ and $^U\text{TM}_{1l}$ resonances within the upper and lower Restrahlen bands are provided as Figs. S13 and S14, respectively. Due to the transverse orientation of the diameter with respect to the propagation direction of the incident photons, the $^U\text{TM}_{1l}$ modes in the upper band appear as near-fields oscillating in phase about the diameter of the structure. This is consistent with expectations for a cavity-based mode. However, the propagating nature of the hyperbolic polaritons is observed within the lower Reststrahlen $^L\text{TM}_{0l}$ modes. In this case the modes depict an oscillating phase to the near-fields, with a clear upward propagation also identified.

**Analytical calculations for spheroidal resonators.** Although there is no analytical solution associated with complex geometries such as the conical nanoresonator shapes described here, analytical results for spheroidal nanoparticles can provide insight. The polariton modes of a spheroid of revolution with semi-axes $a_z$ and $a_x$ can be found by solving the anisotropic Laplace equation for the quasi-static scalar potential. The earliest complete solution of this eigenvalue problem known to us was reported in the early study[18] of magnetostatic waves in ferromagnets. More recently, the solution was independently rediscovered[19] in the context of polariton modes in ZnO quantum dots. The mode spectrum has been shown to satisfy the transcendental equation

$$\sqrt{\varepsilon_z}\sqrt{\varepsilon_x}\frac{d}{d\eta_1}\ln P_l^m(\cosh\eta_1) = \frac{d}{d\eta_0}\ln Q_l^m(\cosh\eta_0), \qquad (4)$$

where $P_l^m(\xi)$ and $Q_l^m(\xi)$ are the associated Legendre functions of the 1$^{st}$ and 2$^{nd}$ kinds and parameters $\eta_0$, $\eta_1$ are defined by

$$\cosh\eta_0 = \frac{a_z}{\sqrt{a_z^2 - a_x^2}}, \cosh\eta_1 = \frac{\kappa a_z}{\sqrt{\kappa^2 a_z^2 - a_x^2}}, \kappa = \frac{\sqrt{\varepsilon_x}}{\sqrt{\varepsilon_z}} \tag{5}$$

For each azimuthal number $m$ and orbital number $l$, Eq. (4) has several solutions that can in principle be labeled by another modal number $r = 1, 2,...$ This number corresponds to the number of radial oscillations in the scalar potential inside the spheroid. Following Walker,[18] one can show that in the case of hBN for non-zero $m$, the number of different radial modes is, respectively, $p + 1$ and $p$ for the upper and lower Reststrahlen band. Here $p$ is the integer part of $(l - m) / 2$. For $m = 0$, these numbers change to $p$ and $l - p$, again for the upper (lower) band. Note that in the case of ZnO the total number of solutions is the same;[19] however, the modes cannot be simply assigned to a particular band because unlike in hBN the two Reststrahlen bands of ZnO overlap spectrally.

We solved Eq. (4) numerically using the standard library root-finding algorithms. The recurrence formulas for Legendre functions were used to compute these functions efficiently. From such calculations we concluded that the best match to the experiment is obtained by assuming that all the observed modes belong to the $r = 1$ family. Indeed, such modes have the largest dipole moment and so the largest coupling to the far-field radiation. The results of these calculations along with the corresponding resonant frequencies of the experimental results where available are provided in Fig. 3 of the main text for the **a** $^U\text{TM}_{1l}$, **b** $^L\text{TM}_{0l}$ modes. Good agreement for the weaker $^L\text{TM}_{1l}$ modes was also found.

**FIGURE CAPTIONS**

**Fig. S1: Reflection Spectra of hBN.** Experimental (blue curve) FTIR reflectance spectra for a 900 nm thick hBN flake on a 500 μm thick BaF$_2$ substrate at 25° incident angle. The corresponding calculated spectra using the transfer matrix approach (red curve) and the best fit optical constants is also provided.

**Fig. S2: Derived Dielectric Function of hBN. a,** Real and **b**, imaginary part of the permittivity for the ordinary ($\varepsilon_t$ green curves) and extraordinary ($\varepsilon_z$ magenta curves) principal axes derived from the best fits of the reflection spectra in Fig. S1. The upper and lower Reststrahlen bands are denoted by the shaded regions in both figures.

**Fig. S3: Isotropic and Hyperbolic Isofrequency Contours.** Schematic representations of the isofrequency curves in wavevector space for an **a**, isotropic medium and a **b**, Type I and **c**, II hyperbolic medium. The green arrow corresponds to the tangent of the isofrequency surface and thus defines the vector of the group velocity.

**Fig. S4: Thickness dependence of hBN reflection. a**, Experimental and **b**, calculated FTIR reflection spectra of hBN flakes of thicknesses ranging from 60 – 6400 nm. The arrows denote the trend in the high frequency edge of the upper Reststrahlen band with reducing thickness while the dashed black lines denote the spectral positions of the LO and TO phonons.

**Fig. S5: Determining hBN nanoresonator geometry.** Scanning electron microscope image of a representative array of hBN nanoresonators. Two resonators detached from the surface during fabrication and provided a method of measuring the top and bottom diameters and the sidewall angle of the structure directly.

**Fig. S6: Geometrical effects upon reflection spectra of hBN nanoresonator arrays.** FTIR reflection spectra of several hBN nanoresonator arrays on SiO$_2$ as a function of **a**, diameter and **b**, thickness (height) for the $^U\text{TM}_{1l}$ modes.

**Fig. S7: Diameter and height dependence of HP resonances.** Spectral position of the $^U\text{TM}_{11}$ (solid symbols, upper half of plots) and $^L\text{TM}_{01}$ (open symbols, lower half of plots) resonances as a function of **a**, diameter and **b,** height for the arrays studied in this work. The lines are provided as a guide to the eye.

**Fig. S8: Transmission and reflection spectra of hBN nanoresonators.** FTIR reflection (red curve) and transmission (blue curve) of a periodic array of 375 nm diameter, 360 nm tall hBN conical nanoresonators fabricated on a 500 μm thick intrinsic silicon substrate. In both cases, the spectra were taken in reference to the surrounding silicon, thus accounting for the >100% reflection.

**Fig. S9: Quality factors and resonant frequencies of hBN nanoresonators. a,** Quality factor and **b**, resonant frequency for the first six $^U\text{TM}_{1l}$ (closed symbols) and first four $^L\text{TM}_{0l}$ resonances (open symbols). The values are reported for two samples, a periodic array of 375 nm diameter hBN nanoresonators on quartz (blue symbols) and on intrinsic silicon (red sybols). The lines are a guide-to-the-eye.

**Fig. S10: Simulated reflection spectra of hBN nanoresonators.** Calculated reflection spectra for a periodic array of hBN nanoresonators at normal (red curves) and grazing (70°, blue curves) incidence for the **a**, lower and **b**, upper Reststrahlen bands. The spectra clearly demonstrate the presence of four distinct series of resonant modes (two per Reststrahlen band) with orthogonal polarization selection rules. For clarity, the normal incidence spectra within the lower Reststrahlen band is amplified by 5x. The modal assignments are also provided. A schematic of the hBN layered structure and the corresponding angles of incidence are provided as an inset in **b**.

**Fig. S11: Magnetic and electric near-fields of hBN nanoresonators in the lower Reststrahlen band.** Simulated spatial distributions of **a**, $|H|$ and **b**, $|E|$ near-fields for the hBN nanoresonators for the first three orders of the $^L\text{TM}_{0l}$ resonances. The modal order $l$ is defined by the number of field oscillations in $|H|$ and the number of field crossing lines in $|E|$ which increase with increasing modal order.

**Fig. S12: Magnetic and electric near-fields of hBN nanoresonators in the upper Reststrahlen band.** Simulated spatial distributions of **a**, $|H|$ and **b**, $|E|$ near-fields for the hBN nanoresonators for the first five orders of the $^U\text{TM}_{1l}$ resonances. The arrows denote the **a**, displacement current, $D$ and **b**, H providing insight into the oscillating nature of the near-fields.

**Fig. S13: Time-variant near-fields within the lower Reststrahlen band.** Simulations of the time-variant **a**, $H_x$ and **b**, $E_z$ near fields for $^L\text{TM}_{03}$.

**Fig. S14: Time-variant near-fields within the upper Reststrahlen band.** Simulations of the time-variant **a**, $H_z$ and **b**, $E_x$ near fields for $^U\text{TM}_{13}$.

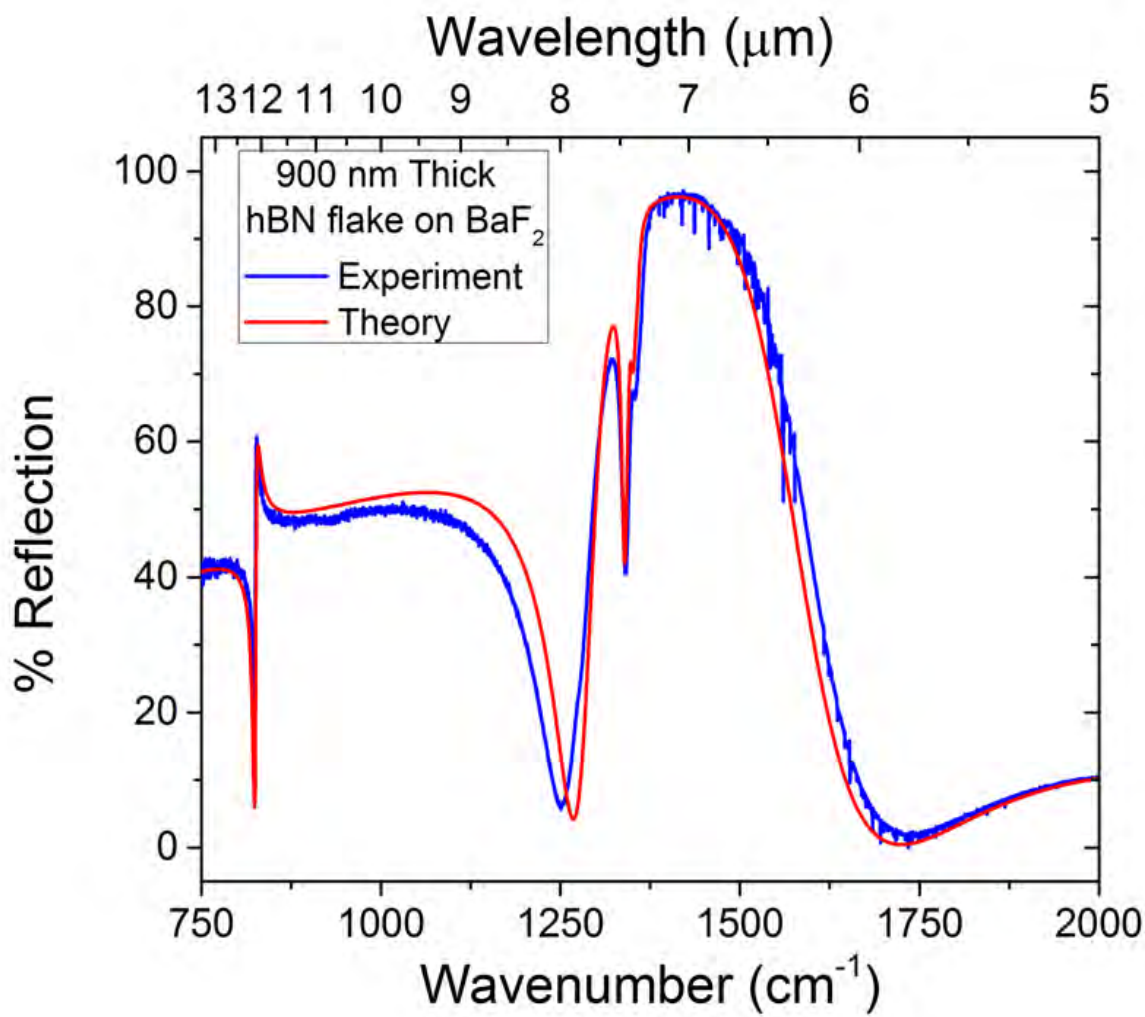

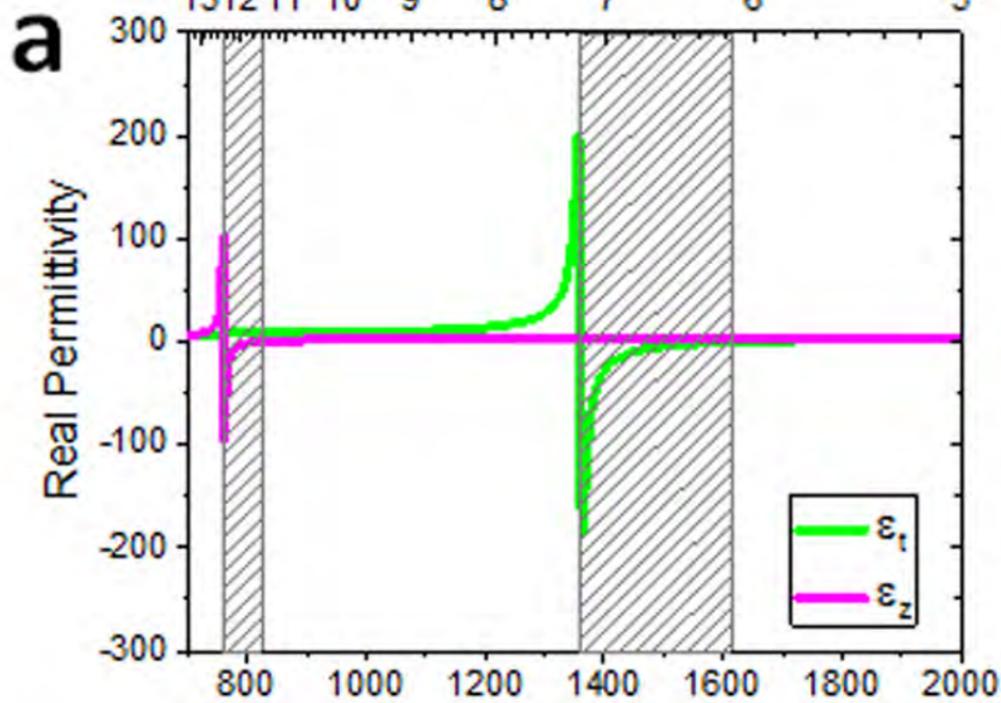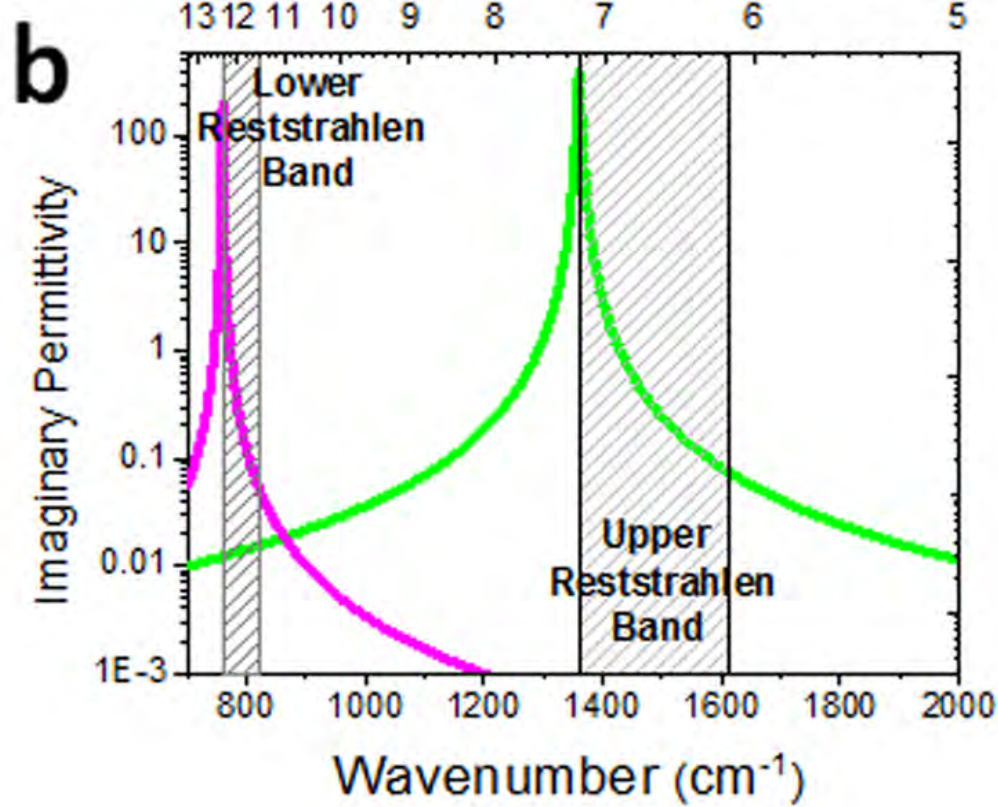

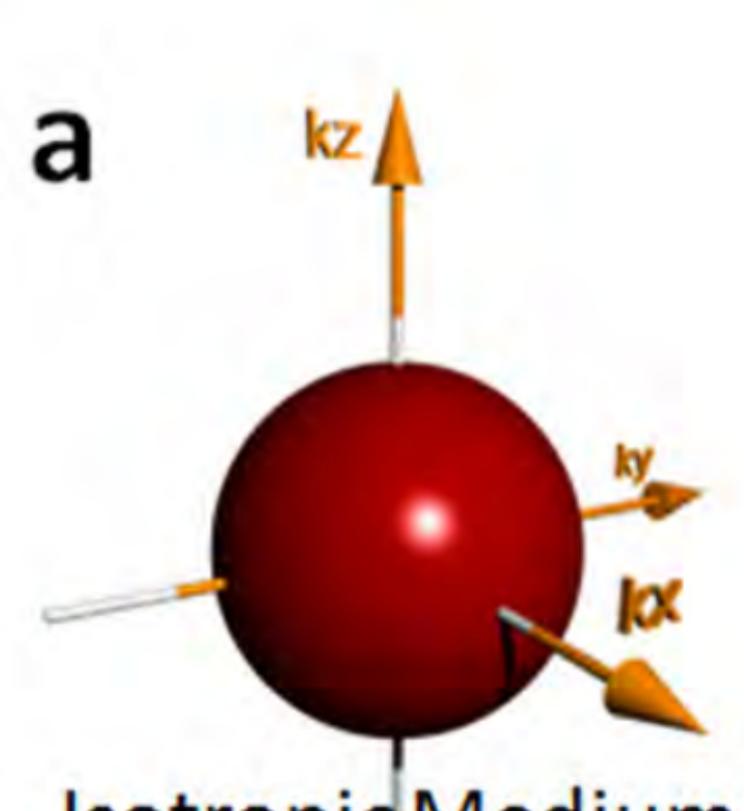 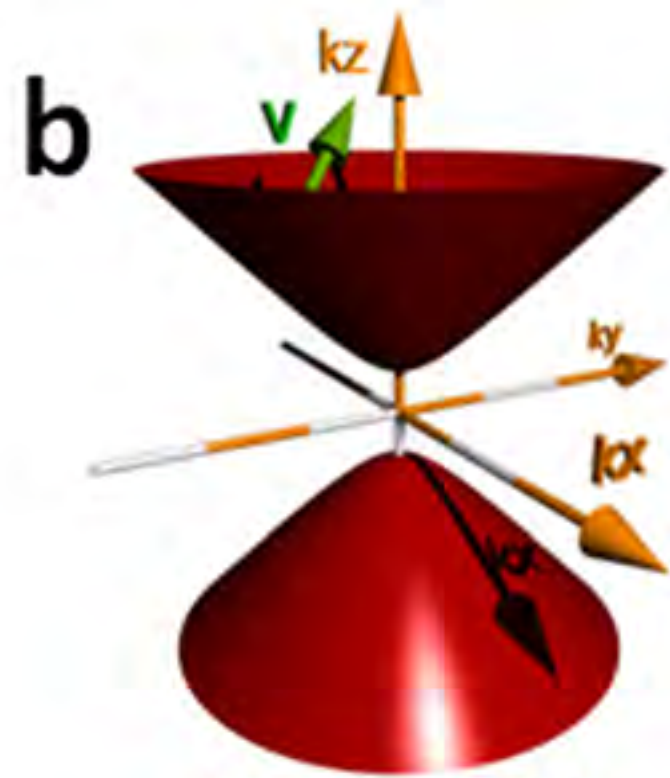 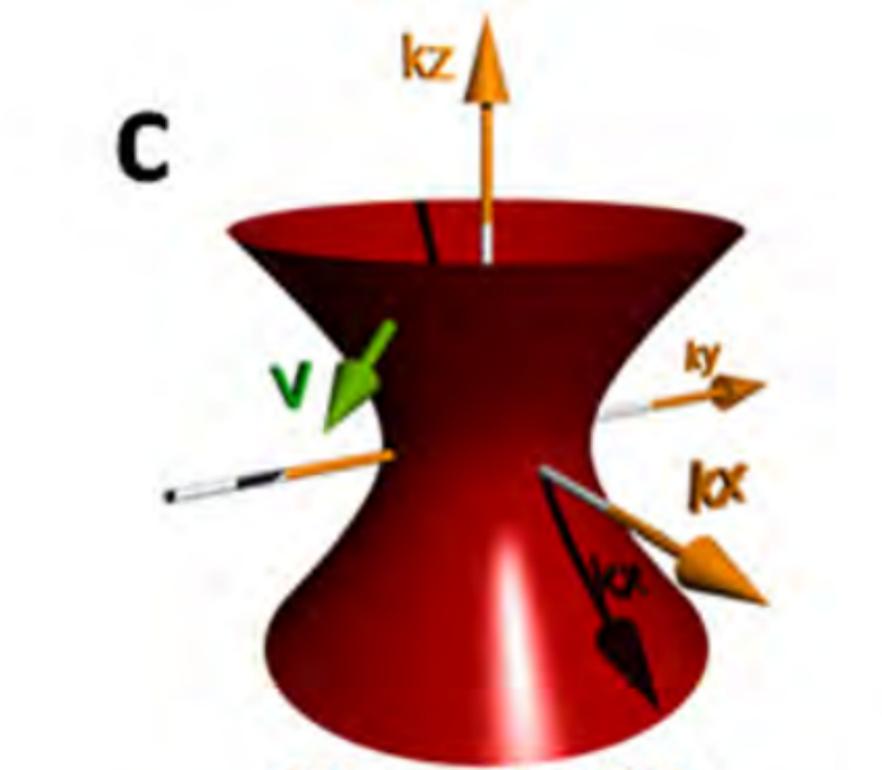

**a** Isotropic Medium $\varepsilon_x = \varepsilon_y = \varepsilon_z$

**b** Type I HM $\varepsilon_x = \varepsilon_y > 0; \varepsilon_z < 0$

**c** Type II HM $\varepsilon_x = \varepsilon_y < 0; \varepsilon_z > 0$

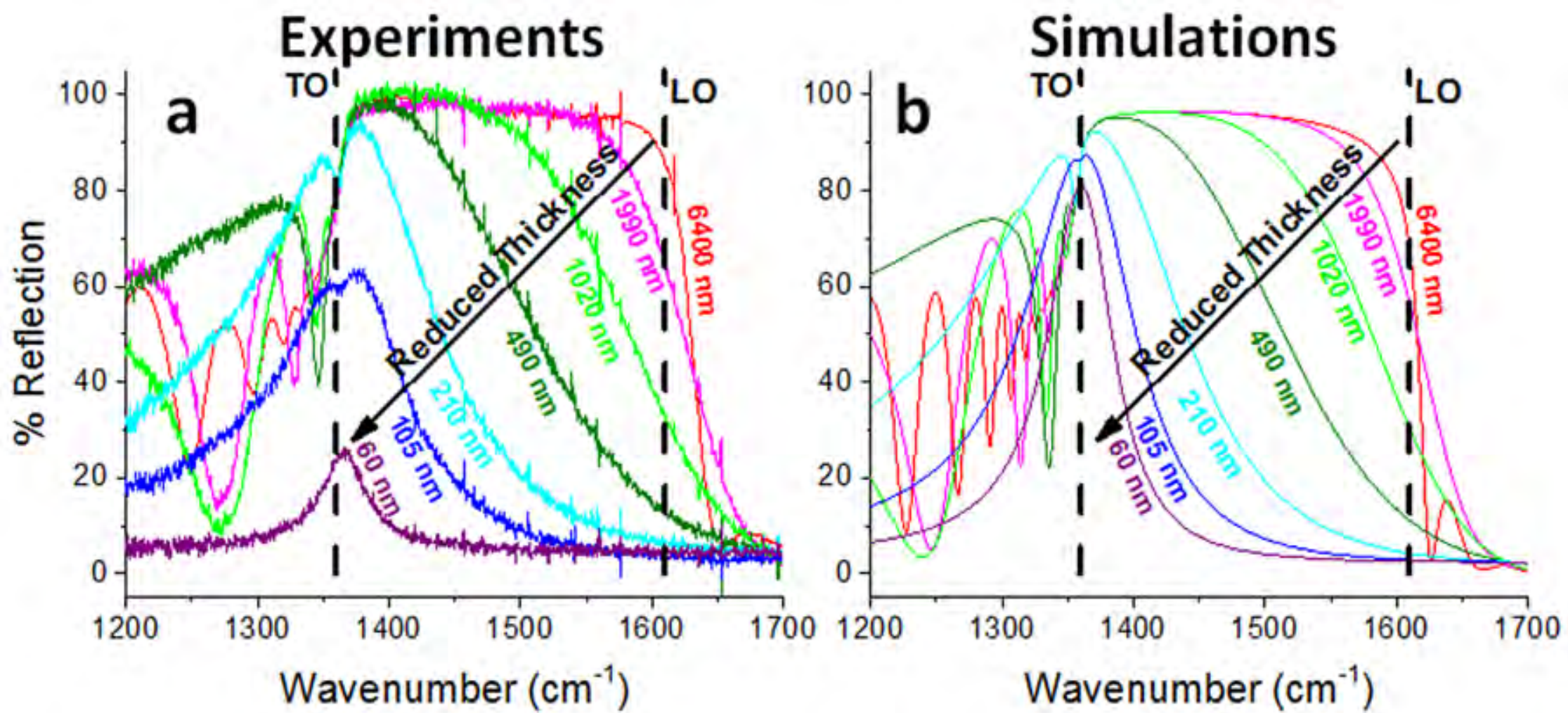

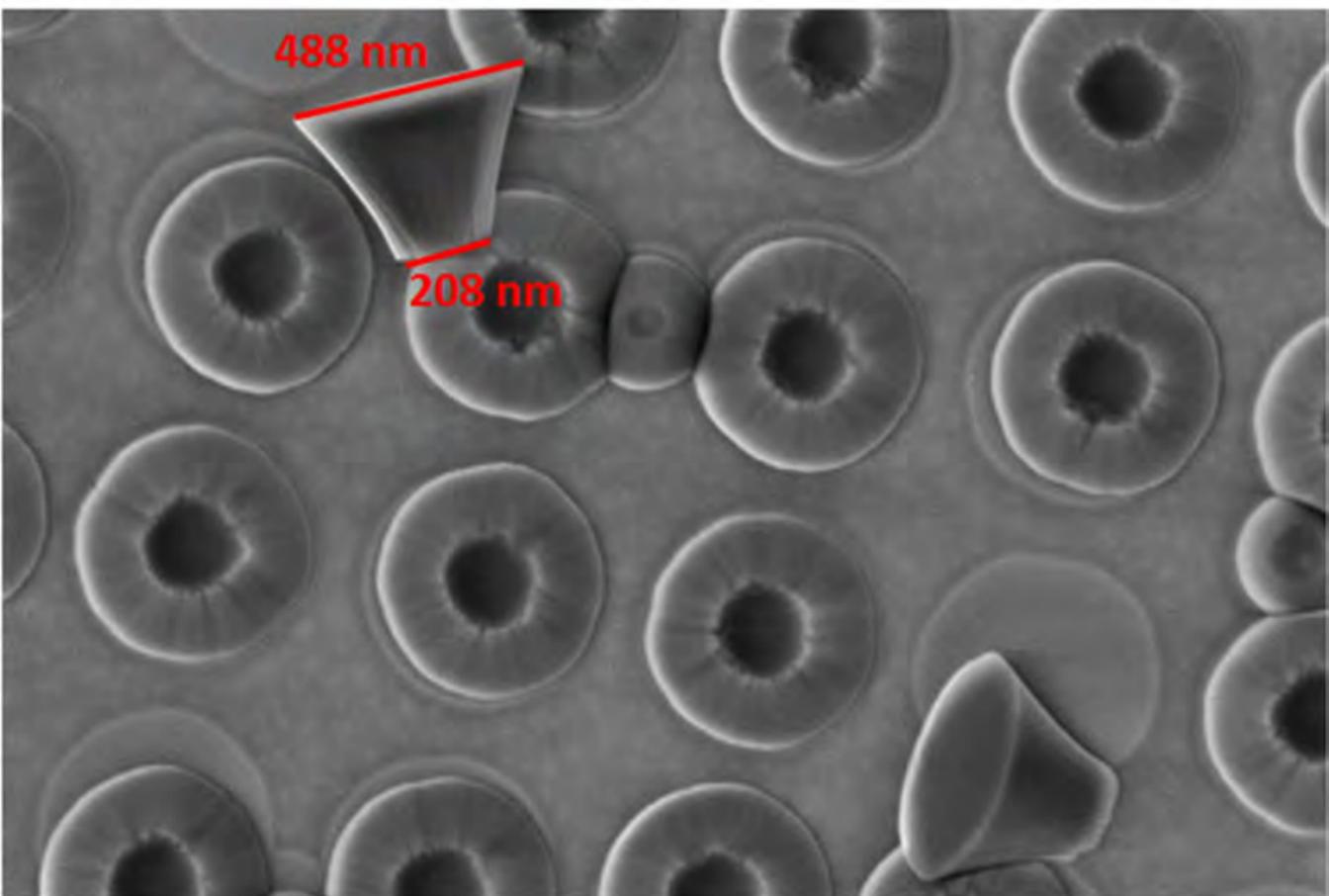

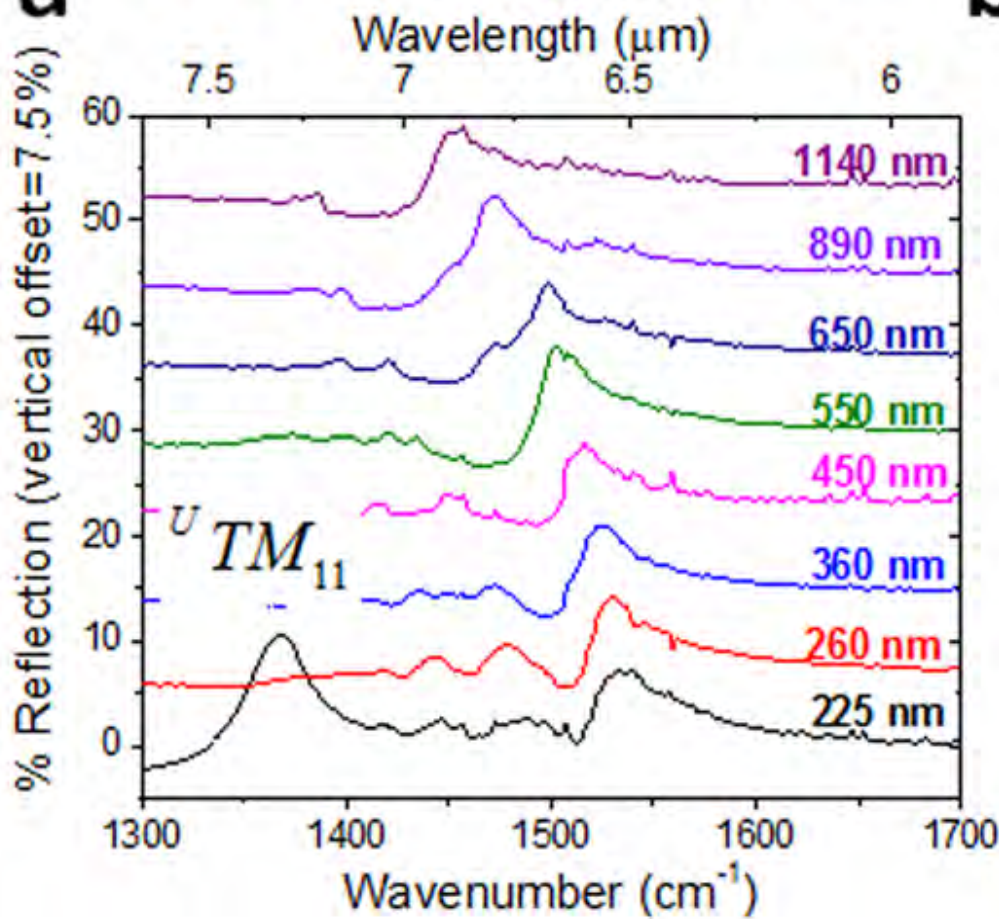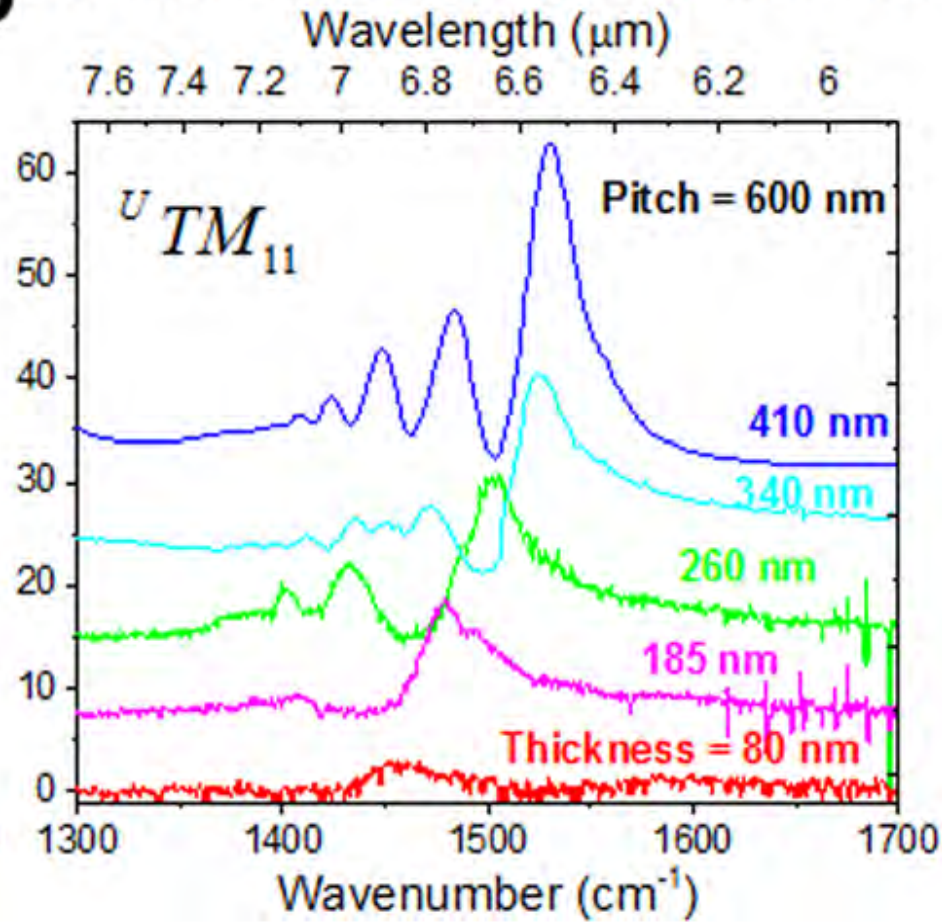

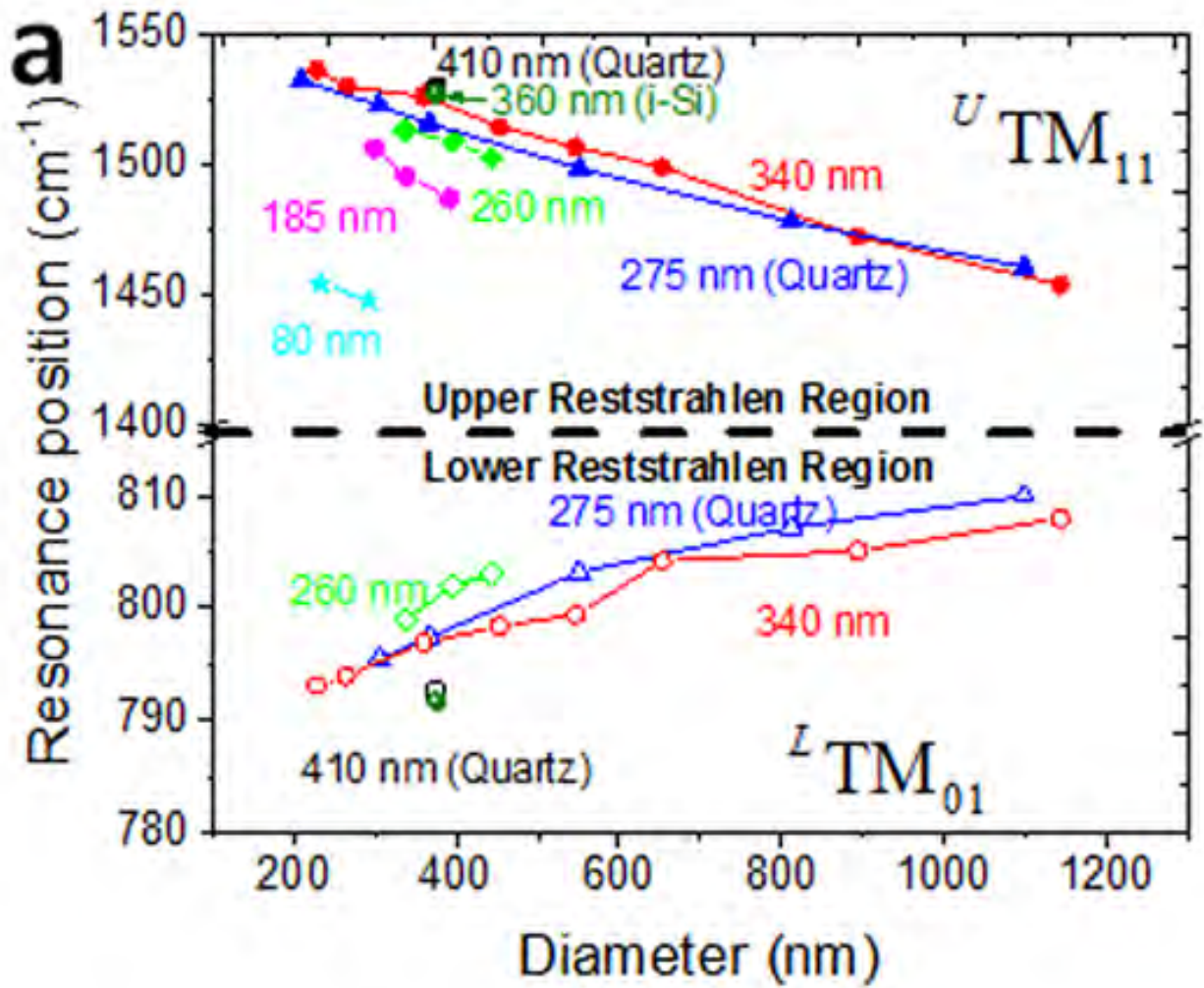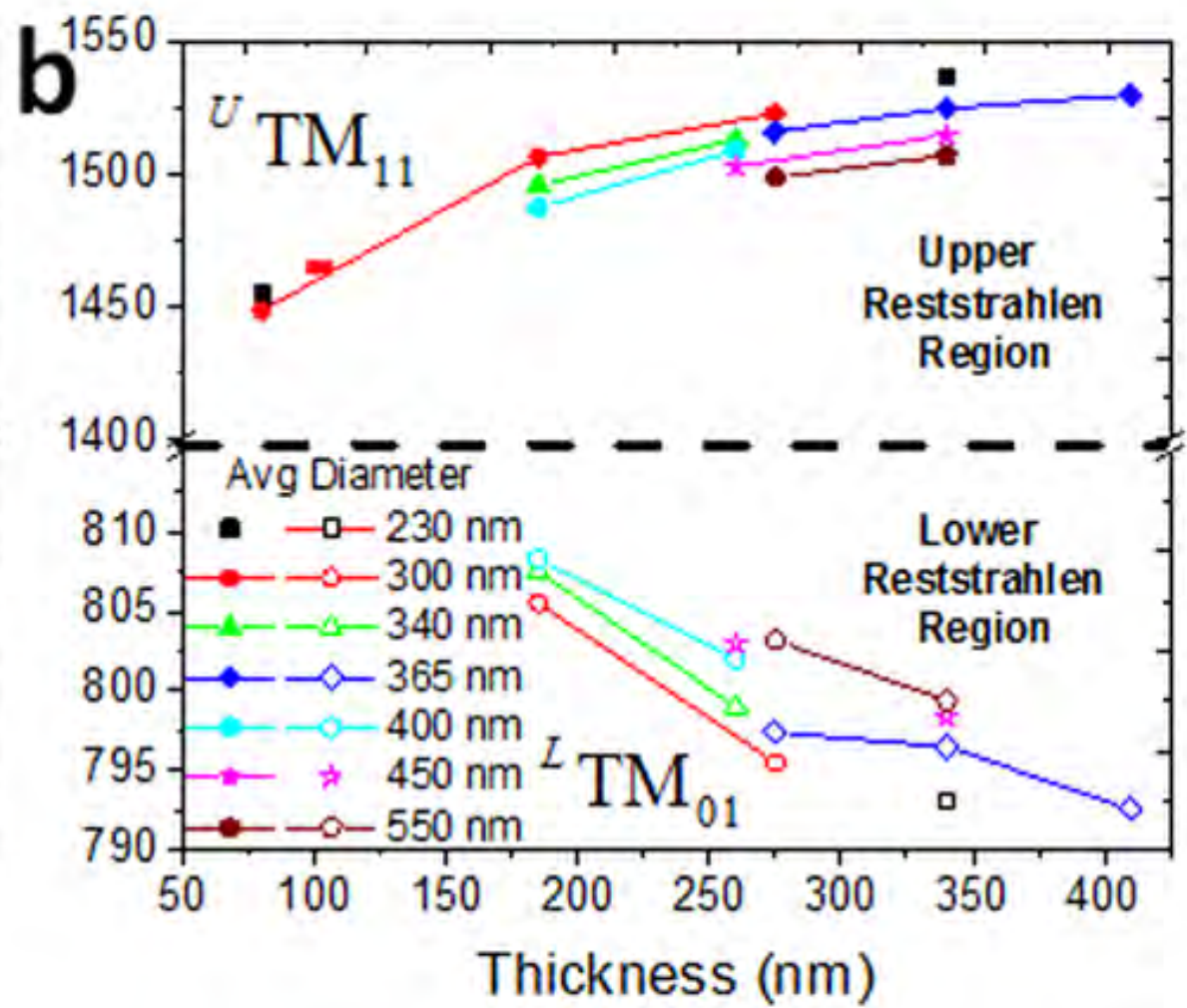

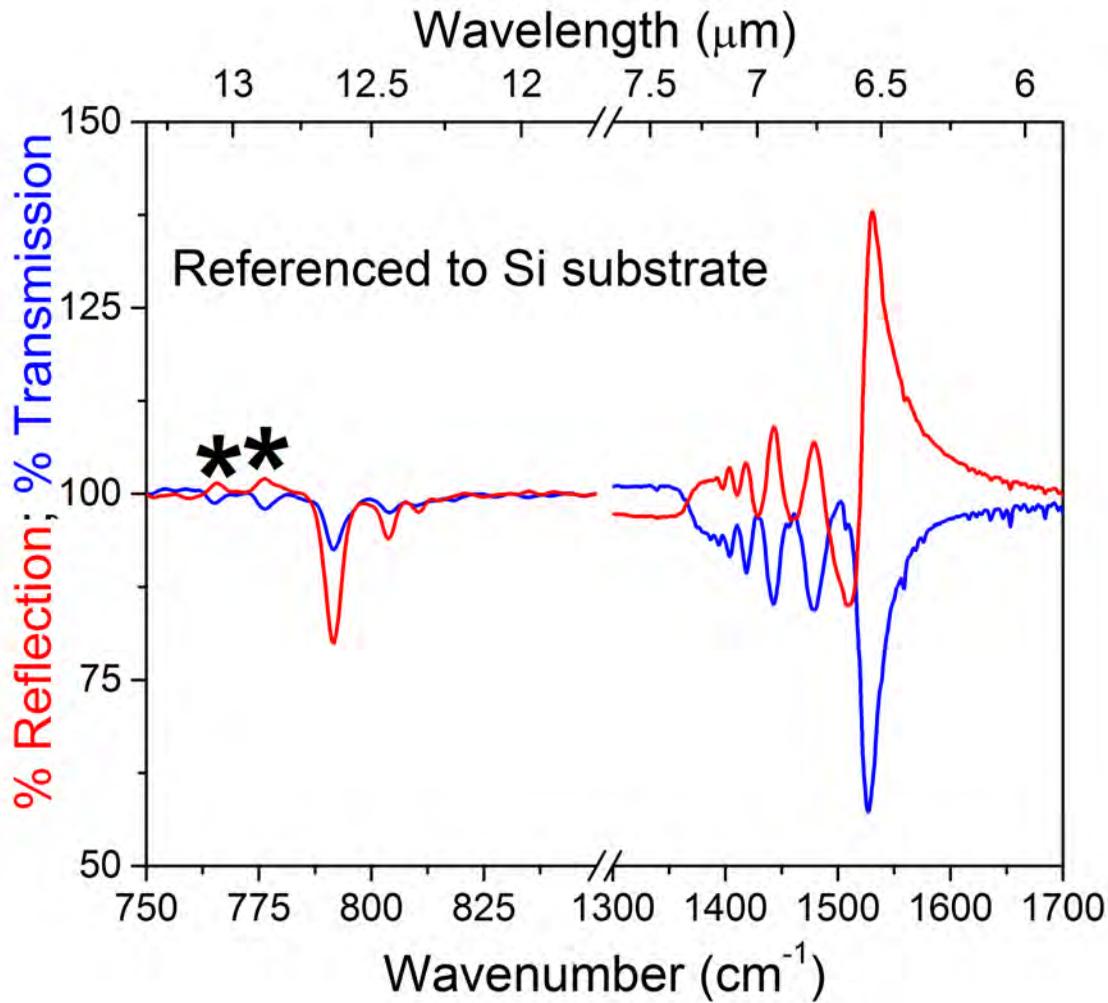

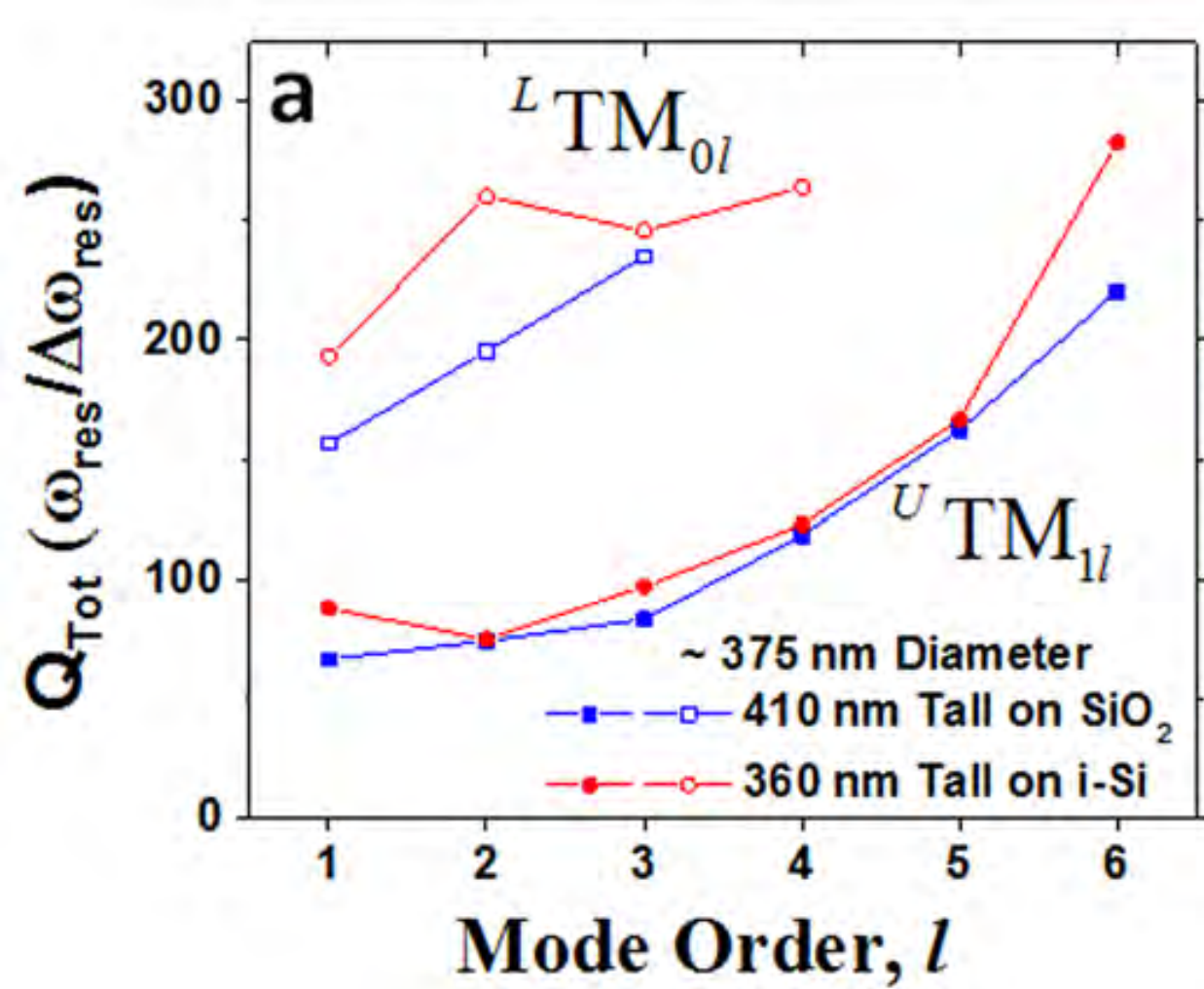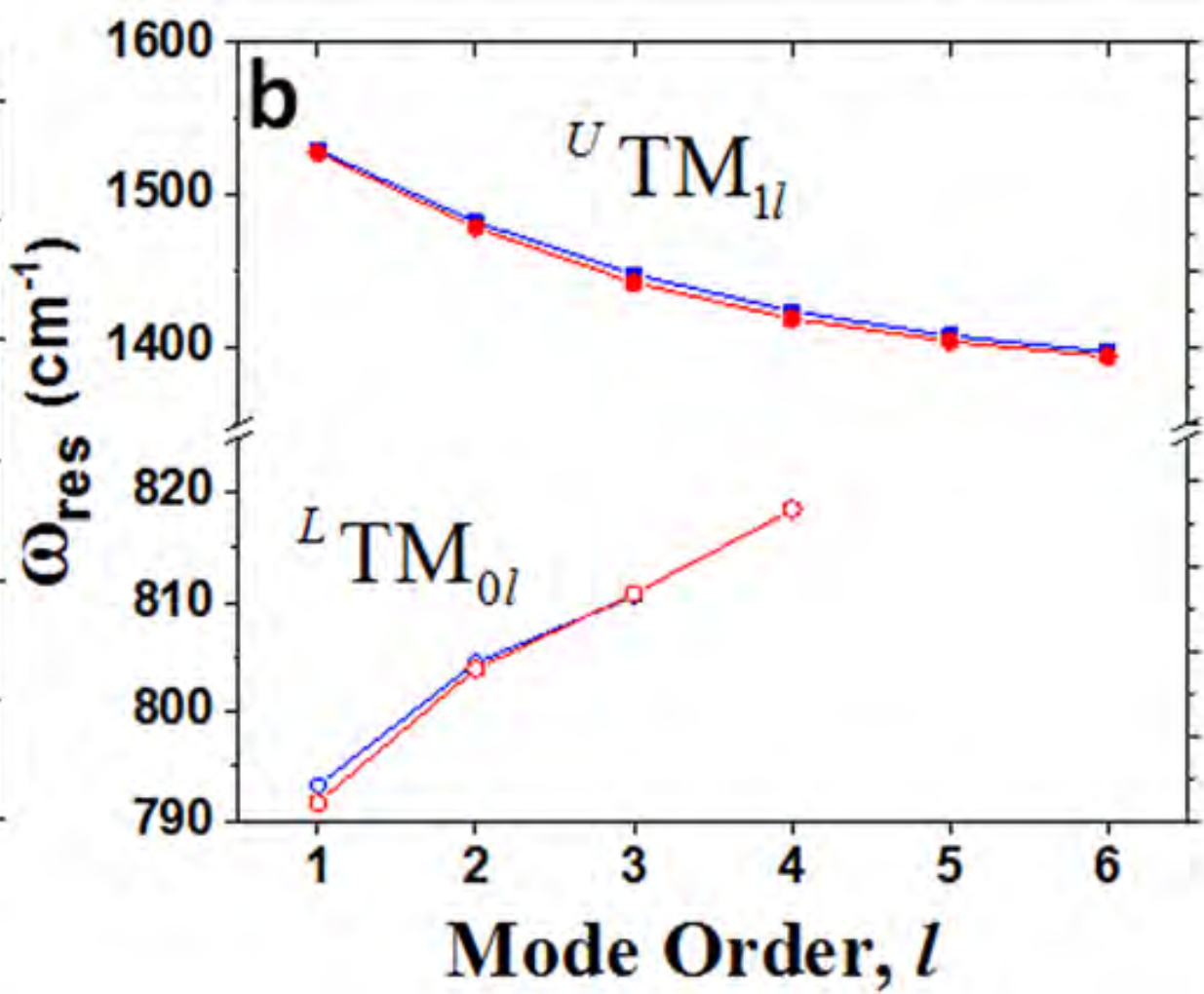

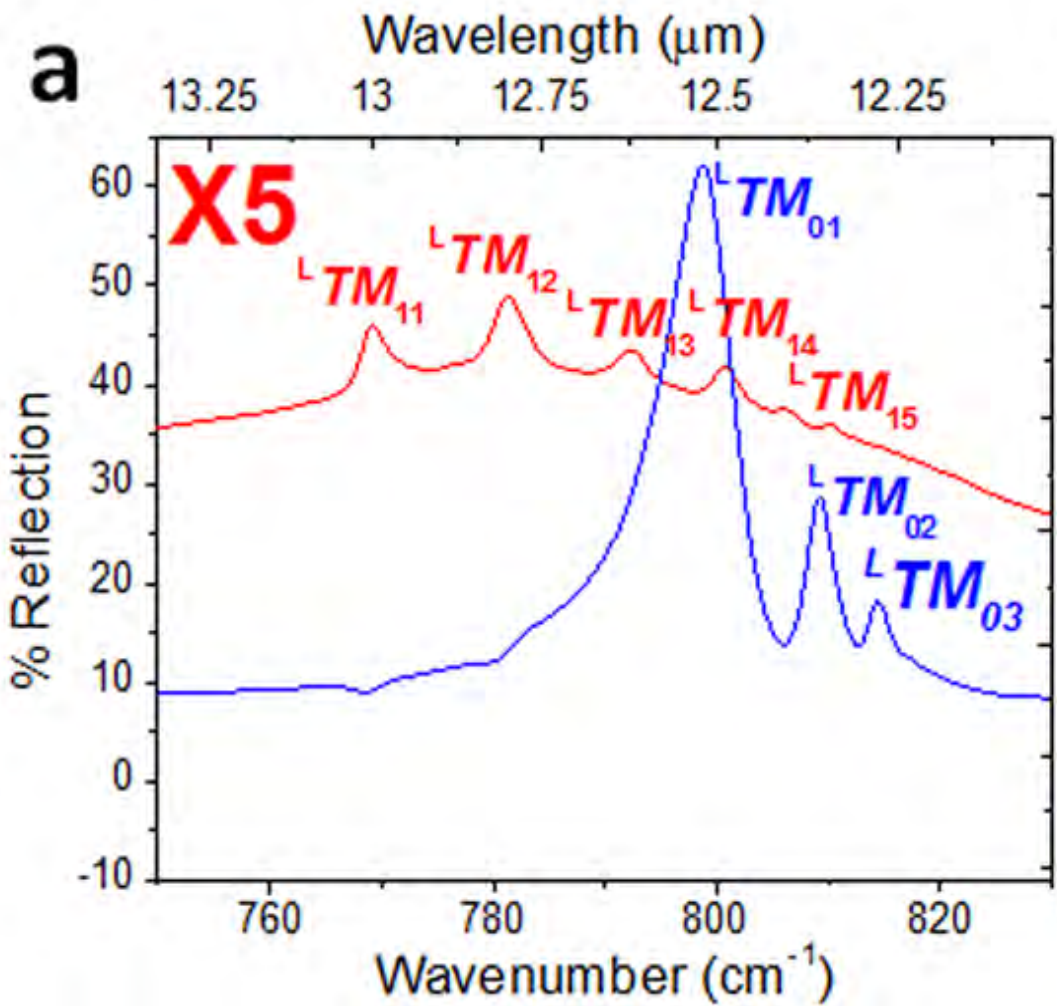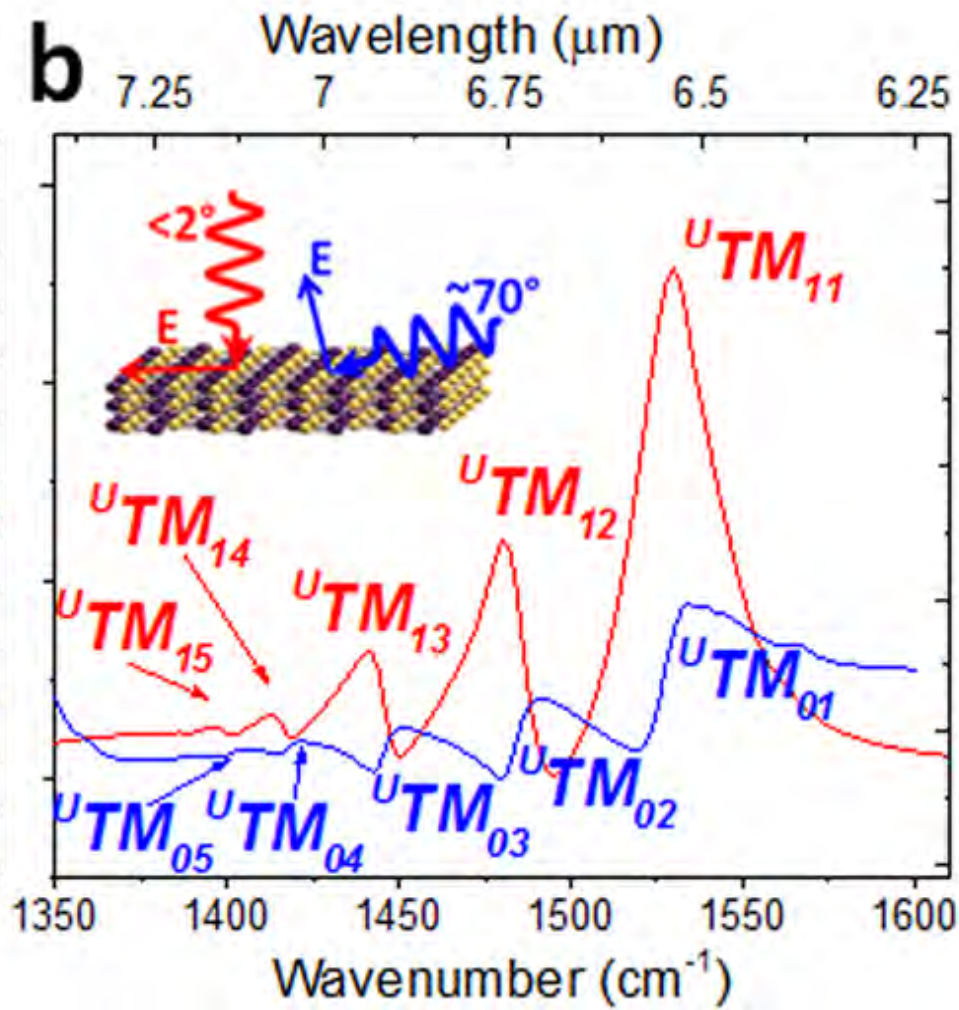

## Lower band

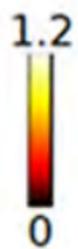 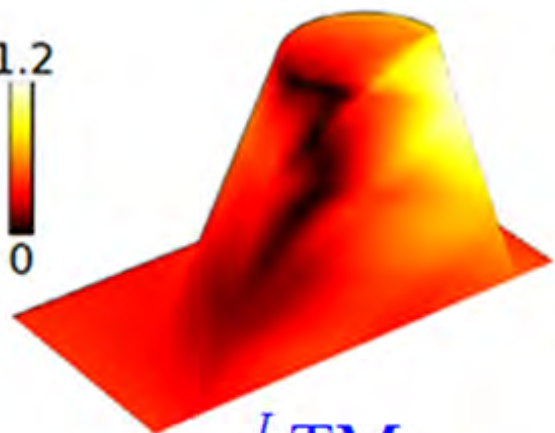 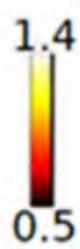 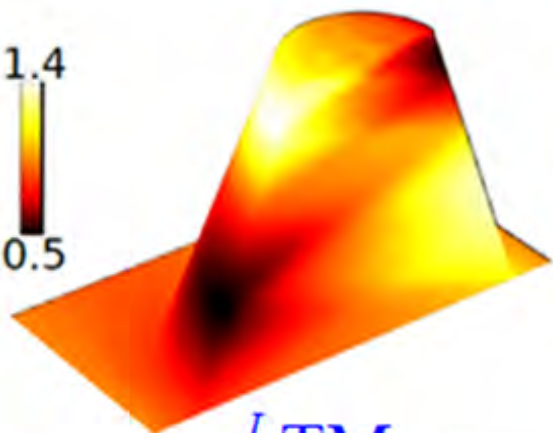 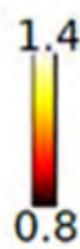 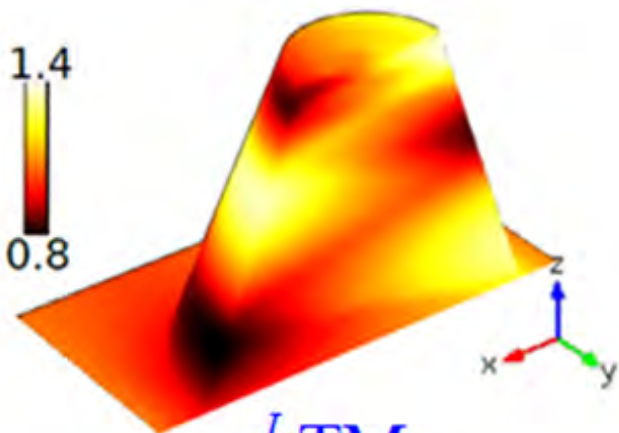 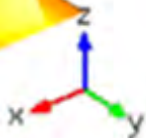

Color: $|H|$

$^L\mathrm{TM}_{01}$  $^L\mathrm{TM}_{02}$  $^L\mathrm{TM}_{03}$

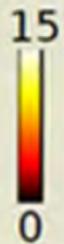 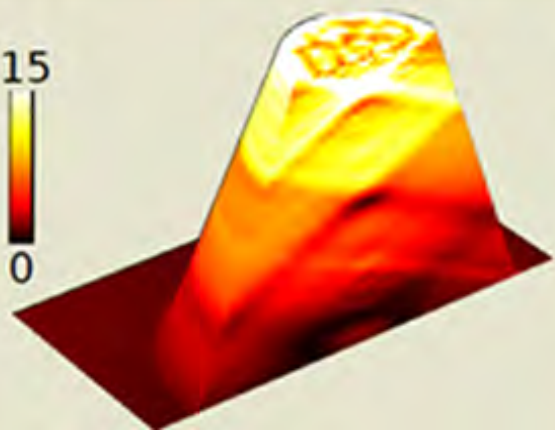 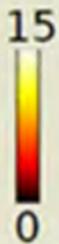 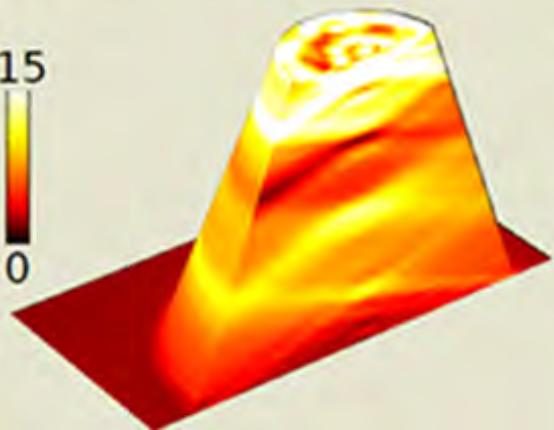 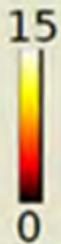 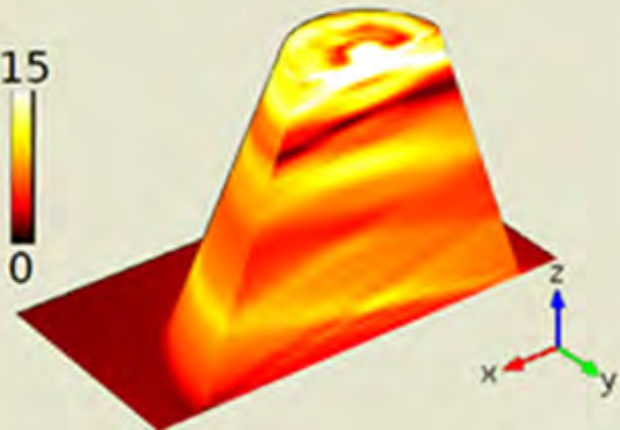 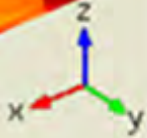

Color: $|E|$

**Upper band**

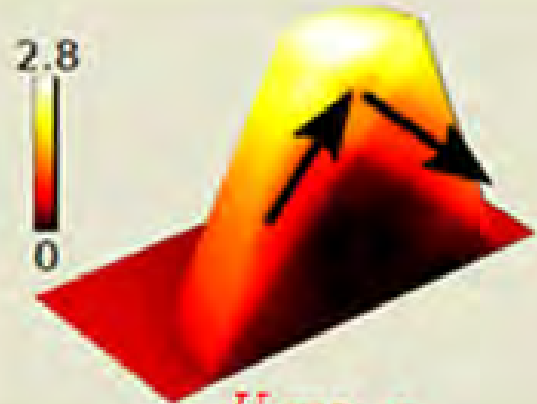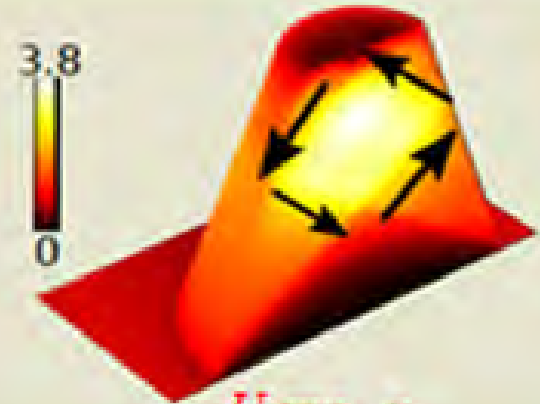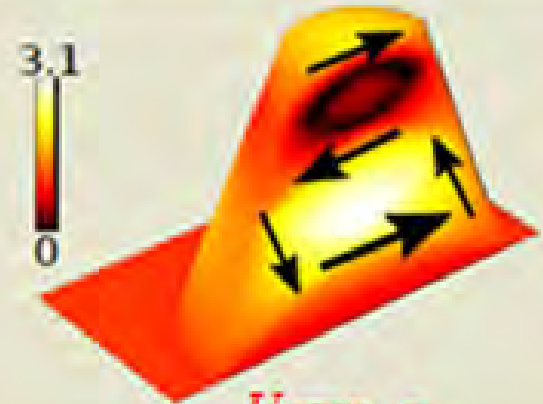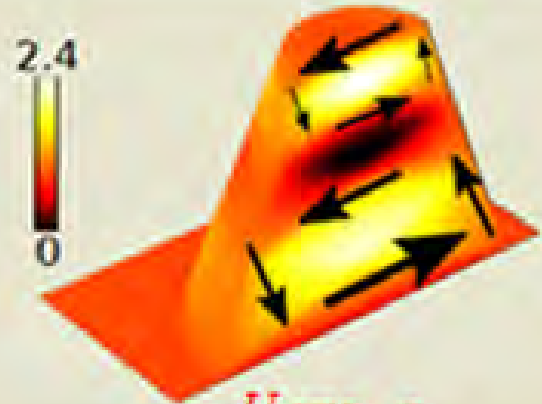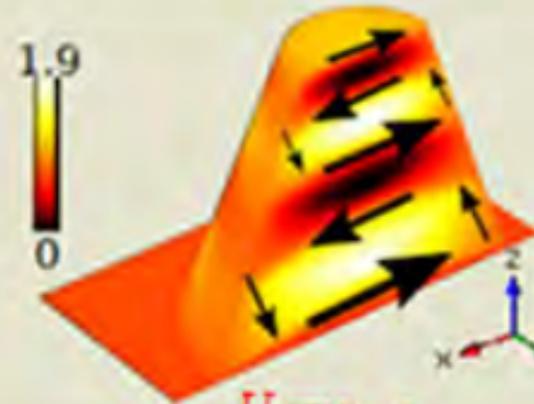

Color: $|H|$
Arrow: $D$

$^U\mathrm{TM}_{11}$  $^U\mathrm{TM}_{12}$  $^U\mathrm{TM}_{13}$  $^U\mathrm{TM}_{14}$  $^U\mathrm{TM}_{15}$

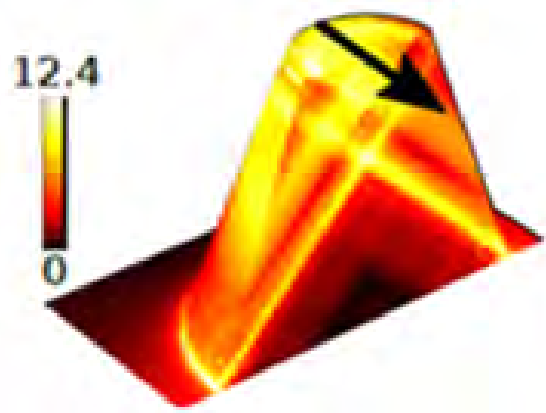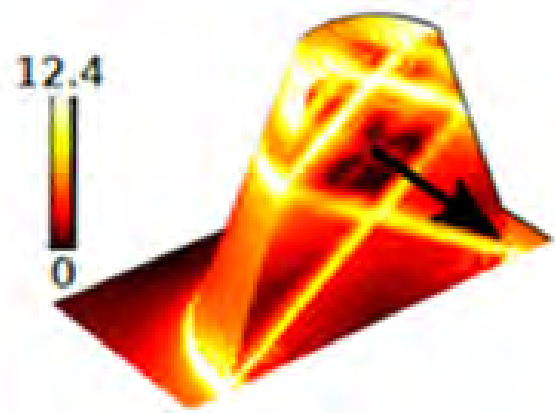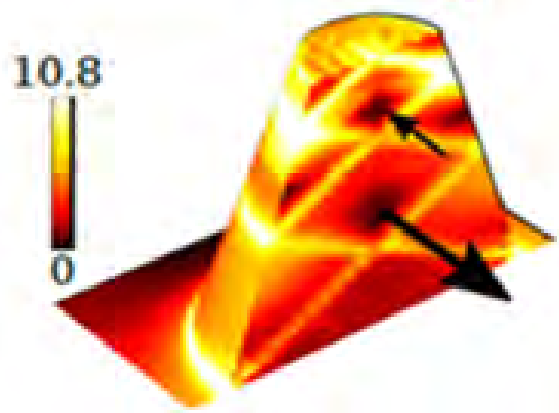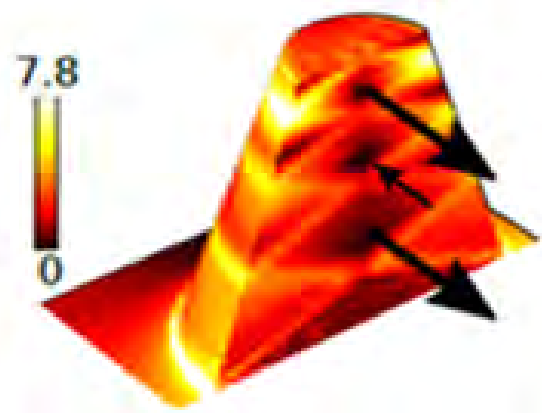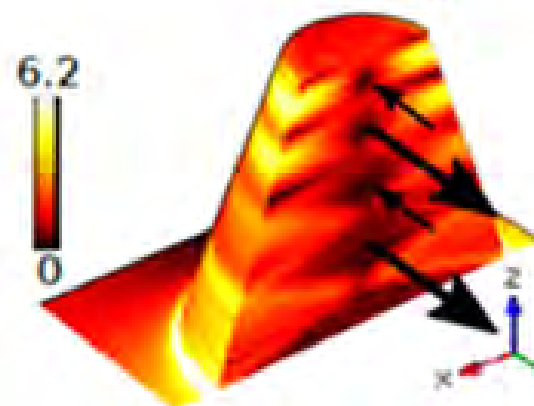

Color: $|E|$
Arrow: $H$